\documentclass[review]{elsarticle}

\usepackage{lineno}
\usepackage{times}
\usepackage{calc}
\usepackage{graphics}
\usepackage{epsfig}
\usepackage{amssymb}
\usepackage{amsmath}
\usepackage{flafter}
\usepackage{float}
\usepackage{latexsym}
\usepackage{psfrag}
\usepackage{dsfont}
\usepackage{rotating}
\usepackage{longtable}
\usepackage{subfigure}
\usepackage{color}
\modulolinenumbers[5]

\journal{Journal of \LaTeX\ Templates}
\renewcommand{\vec}[1]{\mathbf{#1}}

\newcommand{\blue}[1]{\textcolor{black}{#1 }}
\newcommand{\red}[1]{\textcolor{black}{#1 }}

\begin{document}

\begin{frontmatter}

\title{A large eddy simulation method for DGSEM using non-linearly optimized relaxation filters}

\author{David Flad\fnref{myfootnote}}
\author{Andrea Beck}
\address{Institute for Aerodynamics and Gas Dynamics, University of Stuttgart, Pfaffenwaldring 21, 70569 Stuttgart, Germany}
\author{Philipp Guthke}
\address{IWS, University of Stuttgart (now at TransnetBW)}
\fntext[myfootnote]{Corresponding Author}





\begin{abstract}
In this paper, we apply a specifically designed dissipative spatial filter as sub-grid scale model within the increasingly popular discontinuous Galerkin methods and the closely related flux reconstruction high order methods for large eddy simulation.  
The parameters of the filter kernel are optimized with data obtained from direct numerical simulation, that is filtered and used as a ground truth to fit the overall kinetic energy and dissipation rate over time. The optimization is carried out for polynomial degree $3$ to $10$. The optimal kernels are rigorously tested in the limit of infinite Reynolds number flows (HIT and Taylor Green Vortex flow). Additionally, a brief extension to plane turbulent channel flow is given.\\

\end{abstract}


\end{frontmatter}

\section{Introduction}
\label{sec:intro}
Using discontinuous Galerkin (DG) methods and the closely related flux reconstruction (FR) methods \blue{\cite{mengaldo2016}} for large eddy simulation (LES) has become increasingly popular recently. This popularity mainly stems from the high computational efficiency of the method on modern highly parallel super computing clusters combined with geometrical flexibility provided by unstructured grids. Successful LES are reported e.g. in \cite{uranga2011,wiart2012,les1,les2,laf,wiart2015,moura2016}. 
Up to now, it is common practice to use the method without any sub-grid scale models to account for the inevitable effects of under-resolution in LES. This approach is referred to differently as implicit LES (iLES), no-model LES or under-resolved direct numerical simulation (uDNS). \red{ In order to distinguish between true iLES approaches, where the numerical scheme is modified to mimic SGS-model behavior we refer to these schemes as no-model LES in this paper. }All these methods use the numerical viscosity introduced by the numerical flux function (for both viscous and advective fluxes) that couples the inter-cell discontinuities as a surrogate for turbulence closure. Recently it was shown in \cite{flad2017} that the applicability of this approach is limited to comparably well resolved LES, where the molecular dissipation resolved on the coarse grid accounts for at least $\approx 45\%$ of the total dissipation resolved by direct numerical simulation (DNS). For typical coarse LES, especially considering high Reynolds numbers and realistic, engineering application scenarios, the resolved dissipation drops below $1\%$ rendering the above methods inaccurate. In \cite{flad2017} a method was proposed to use a novel form of the DG operator \cite{gassner2016split} that guarantees consistency in the discretization of kinetic energy as well as preservation (called KEP-flux) in combination with a low dissipative Riemann solver (so called $RoeL^2$) 
\cite{osswald2015l2roe} and a simple explicit Smagorinsky model \cite{smago}. It was shown that by using the KEP-flux, the LES is stable without further need for de-aliasing or even interface dissipation, giving full freedom to shape the introduced numerical viscosity based upon turbulence modeling considerations. Thereby, numerical stability constraints and modeling considerations are decoupled. With the proposed modeling, excellent results were obtained for typical coarse LES, while the low dissipation Riemann solver suppressed density and pressure oscillation arising due to interface discontinuities. The method was further extended for transitional flows by using a high pass filtered variant of Smagorinskys model in \cite{manzanero2018}.\\
In this work the basic properties of KEP-flux described above are exploited again to design a filter-based LES scheme, tailored specifically for the underlying numerical method. As a baseline, the KEP-flux is used with the $RoeL^2$ Riemann flux at interfaces. Then modal filter kernels are searched by non-linear optimization using the well known Nelder-Mead algorithm, with the objective to minimize the mean square error of integral kinetic energy and resolved dissipation over time. The optimization is run for the test case of decaying isotropic turbulence already used in 
\cite{flad2017} ($Re_{\lambda}\approx162$ to $Re_{\lambda}\approx97$). Parameters of the optimization are the diagonal coefficients of a modal filter matrix as well as a factor scaling the filter strength. The filter strength, motivated by the spectral eddy viscosity method (SPEVM)\cite{chollet1985}, is scaled by the kinetic energy content of the highest polynomial mode. In contrast to most explicit turbulence closures for LES, the proposed filtering approach does not require any computation of second order derivatives. The computation of second order derivatives typically accounts for more than $50\%$ of the computational cost of the DG operator, while the effect of the resolved molecular dissipation for high Reynolds, coarse grid LES is often negligible.\\
The optimization is shown to give excellent results for polynomials of degree $N=3$ to $N=10$. Afterwards the method is tested for infinite Reynolds number decaying isotropic turbulence with a Kolmogorov spectrum and the inviscid Taylor-Green-Vortex flow.

\section{Numerical Methods}
\label{sec:numerical_method}

\subsection{Discontinuous Galerkin Spectral Element Method}
\label{sec:dgsem}

We consider the compressible Navier-Stokes equations (NSE) expressed in conservation form 
\begin{equation}\label{eq:NSE}
\vec{U}_t + \vec{\nabla}_{x}\cdot\vec{F}(\vec{U},\vec{\nabla}_x \vec{U}) = 0,
\end{equation}
where $\vec{U}$ denotes the vector of conserved quantities $\vec{U}=(\rho, \rho u, \rho
v, \rho w, \rho e)^T$, the subscript $t$ the time derivative and $\vec{\nabla}_x$ the gradient operator in physical space. The flux
is the difference of advection and viscous fluxes, $\vec{F}=\vec{F}^{a}(\vec{U})
- \vec{F}^{v}(\vec{U},\vec{\nabla}_{x}\vec{U})$, with the entries \red{
\begin{equation}
\vec{F}^{a}_{l}(\vec{U})=\begin{pmatrix}\rho\, v_l\\ \rho\, u v_l +\delta_{1l}\,p\\
\rho\, v v_l +\delta_{2l}\,p\\ \rho\, w v_l +\delta_{3l}\,p  \\ \rho\,ev_l + p\,v_l\end{pmatrix},\; 
\vec{F}^{v}_{l}(\vec{U},\vec{\nabla}_{x}\vec{U})=\begin{pmatrix}0\\
\tau_{1l}\\
\tau_{2l}\\\tau_{3l}\\ \tau_{lj}v_j - q_l\end{pmatrix},
\end{equation}}
where $l=1,2,3$, denoting the Cartesian directions of the flux $\vec{F}_1,\vec{F}_2,\vec{F}_3$. We follow the usual nomenclature for \red{ $\rho, \vec{v}=
(u,v,w)^T, p,
e$} denoting the density, velocity vector, pressure and specific total energy, respectively.
\red{With the viscousx flux we introduced the viscous stress tensor $\underline{\vec{\tau}}$ and the heat flux $\vec{q}$ 
\begin{equation}\label{eq:viscStress}
\underline{\vec{\tau}}:=\mu (\vec{\nabla} \vec{u} +
(\vec{\nabla}\vec{u})^T -
\frac{2}{3}(\vec{\nabla}\cdot
\vec{u})\vec{\delta}),
\end{equation}
\begin{equation}\label{eq:heatFlux}
\vec{q}=-\lambda\vec{\nabla}T,
\end{equation}
with $\lambda=\frac{c_p \mu}{Pr}$. Fluid dependent variables are heat conductivity $\lambda$, dynamic viscosity $\mu$, Prandtl number $Pr$ and the specific 
heats $k$, 
$c_p$ 
and $c_v$., and are assumed to be constant. Using
the adiabatic coefficient $\kappa=\frac{c_p}{c_v}$ and the specific gas constant
$R=c_p-c_v$, the system of equations is closed by the perfect gas law}
\begin{equation}
p=\rho R T=(\kappa-1)\rho(e-\frac 1 2
\vec{v}\cdot\vec{v}),\quad
e=\frac{1}{2}\vec{v}\cdot\vec{v} + c_v T.
\end{equation}

In this work, we use a special DG variant, namely the discontinuous Galerkin spectral element collocation method (DGSEM) with Legendre Gauss-Lobatto (LGL) nodes. 
The LGL nodes are essential, as this choice guarantees the so-called summation-by-parts (SBP) property of the resulting DGSEM operator 
\cite{gassner_skew_burgers}. Up to now, split form DG is only available for this specific variant, as the SBP property is fundamental. The computational domain is subdivided into non-overlapping hexahedral elements which are transformed to reference space $\vec{\xi}$ via a transfinite mapping. 
Within the reference element, a tensor-product polynomial approximation is constructed: we use the tensor-product of the 1D LGL nodes and the associated tensor-product of 1D Lagrange polynomials, which gives $(N+1)^3$ DOF per element per unknown quantity. The method was described in \cite{flad2017}, stemming from \cite{gassner2016split}. 
For details the interested reader may refer to these original papers.
In this work the split form DG is used. As introduced by \cite{fisher2013,carpenter2014} the \red{cartesian flux derivative after transformation to reference space in split form (exemplary for one reference direction $\xi$) at one LGL node $(i,j,k)$ reads}
\begin{equation}
\begin{split}
\frac{1}{\Delta x} \vec{F}^a_1(\vec{U})_\xi\big|_{ijk} & \approx  \frac{1}{M_{ii}}\left(\delta_{iN}\left[\vec{F}_1^{a,*} - \vec{F}_1^a\right]_{Njk} - 
\delta_{{i0}}\left[\vec{F}_1^{a,*} - \vec{F}^a_1\right]_{0jk}\right)\\
&+\sum_{m=0}^N 2\, D_{im}\vec{F}^{a,\#}_1(\vec{U}_{ijk}, 
\vec{U}_{mjk}),\\
\end{split}
\end{equation}
where $\vec{F}^{a,\#}_1(\vec{U}_{ijk}, \vec{U}_{mjk})$ is a two-point numerical volume flux \red{and $D_ij$, $M_ij$ are the polynomial derivative matrix and the (lumped) mass matrix respectively}. In this work the kinetic energy 
consistent flux as introduced in \cite{gassner2016split} is used, reproducing the well 
known split form of Pirozzoli \cite{pirozzoli2011numerical}
  \begin{equation}
  \label{eq:PIflux}
      \vec{F}^{a,\#}_1(\vec{U}_{ijk},\vec{U}_{mjk})=\begin{bmatrix}
                            \{\{\rho\}\}\{\{u\}\} \\
                            \{\{\rho\}\} \{\{u\}\}^2 + \{\{p\}\} \\
                            \{\{\rho\}\} \{\{u\}\} \{\{v\}\} \\
                            \{\{\rho\}\} \{\{u\}\} \{\{w\}\} \\
                            \{\{\rho\}\} \{\{u\}\} \{\{h\}\}
                            \end{bmatrix},
   \end{equation}
   with
 \begin{equation}\nonumber
    \{\{\alpha\}\} := \frac{1}{2}\,(\alpha_{ijk}+\alpha_{mjk}).
   \end{equation} \red{and the enthalpy $h$ given by $e + p/\rho$.}
In \cite{gassner2016split}, it was also shown that when choosing the element interface flux $\vec{F}_1^{a,*}$ equal to \eqref{eq:PIflux}, the resulting split form DGSEM is kinetic energy preserving across the domain. Effectively, this means that the aliasing error in the kinetic energy preservation due to the discretization of advective terms is eliminated. A Roe type matrix dissipation is added at element interfaces to recover an unwinding type split form DG scheme. For the viscous numerical flux function the dissipation free Bassi and Rebay scheme BR1 is used \cite{Bassi&Rebay:1997:B&F97,bassi2,br1isstable}.
Finally, the semi discrete (split form) DGSEM is integrated in time with an explicit fourth order low storage Runge-Kutta method, \cite{Carpenter-MH:2005kx}. 

\subsection{Modal filter implementation}
 \label{sec:filtering}
The LES method in this work is based upon a specifically optimized filter shape in modal polynomial space. This section describes how to generally 
construct and apply modal filtering directly to nodal polynomial representations. \red{We describe the filtering in 1D for clarity, the extension to 2D/3D follows by tensor products extension (line by line).} For filtering we need the 
modal orthogonal Legendre basis ${\varphi_j(\xi)}^N_{j=0}$ evaluated at the interpolation points $\xi_i$. The Legendre basis is given by the recursion formula:
\begin{equation}
 \varphi_{j+1}(\xi) = \frac{2j+1}{j+1}\xi\varphi_{j}(\xi) - \frac{j}{j+1}\varphi_{j-1}(\xi),\quad j=1,...,N-1
\end{equation}
with the starting point $\varphi_0(\xi)=1$, $\varphi_1(\xi)=\xi$. The Legendre basis functions are then normalized by $\sqrt{j+0.5}$ so that $\varphi_j(1)=1$. 
The 1D Vandermonde matrix is defined by $V_{ij}:=\varphi_j(\xi_i)$, $i,j = 0,...,N$. Thereby the \red{1D solution vector of a scalar quantity in nodal space $\vec{u}$} is transformed to modal space by \red{multiplying with} the inverse \red{of the 1D Vandermonde}:
\red{
\begin{equation}
 \tilde{\vec{u}} = \underline{\vec{V}}^{-1} \vec{u}
\end{equation}}
In modal space, the filter matrix is defined by
\red{
\begin{equation}
 \widetilde{K}_{ij} = \delta_{ij} \sigma_i,\quad i,j=0,...,N,
\end{equation}}
with the modal filter coefficients $\sigma_i \in [0,1]$ free to be chosen to design a specific dissipation behavior. Note that fixing $\sigma_0 =1$ results in a conservative filter.
Applying the filter matrix the filtered modal solution is:
\red{
\begin{equation}
 \overline{\tilde{\vec{u}}} = \underline{\tilde{\vec{K}}}\tilde{\vec{u}}.
\end{equation}}
Finally the filtered modal solution vector is transformed back to nodal space and the filtered solution vector in nodal space is:
\red{
\begin{equation}\label{eq:projectionfilter}
 \overline{{\vec{u}}} = \underline{\vec{V}}\overline{\tilde{\vec{u}}} = \underbrace{\underline{\vec{V}} \underline{\vec{K}} \underline{\vec{V}}^{-1}}_{\underline{\vec{K}}} {\vec{u}}
\end{equation}}
The above steps can be efficiently implemented by a single matrix vector multiplication in which the filter matrix \red{$\underline{\vec{K}}$} defined in \eqref{eq:projectionfilter} is applied to
$\vec{u}$.\\
Our filter implementation differs in an important aspect from the usual approach in DG methods: applying the filter directly to the solution vector at each Runge-Kutta (RK) stage results in a dissipative filter effect that is proportional to the inverse of the time step (artificial dissipation added by the filter)~\cite{Hesthaven2008}. This dependence makes the common approach unsuitable for a filter-based LES. In order to obtain a filter whose dissipative action is independent of the time step and can thus serve as a closure model, the filter has to be applied within the time step update as follows:
\begin{equation}
 \vec{u}_{t^*} = \vec{u}_t + \sigma_F (\bar{\vec{u}} - \vec{u}),
\end{equation}
where $\bar{\vec{u}}$ is the filtered solution and $\sigma_F$ is the scalar filter strength. For the first RK stage for example this translates to
\begin{equation}
 \vec{u}^{n+1} = \vec{u}^n + b_{1}(\vec{u}_t + \sigma_F (\bar{\vec{u}} - \vec{u})) 
\end{equation}
where $b_1$ is the first stage RK time step size. Thus, the filter strength $\sigma_F$ is multiplied by the time step size, recovering independence of filter effect and time step size.
Note that the properties of the spatial operator are not affected by this filtering procedure, as the filtering is not applied within it.

\section{Sub-grid scale model: the optimized filter}\label{DGiles}
It has previously been shown \cite{flad2017,manzanero2018} that for typical underresolved LES with DGSEM an explicit LES model in combination with a low Mach Riemann solver gives the best results. In this section a strategy is explained to design a LES model specifically tailored to DGSEM. The aim now is to replace the Smagorinsky model, while using the same baseline scheme (KEP-DG with low Mach/low dissipation Riemann solver $L^2Roe$). The advantage of DGSEM, compared e.g. to finite volume methods, is its inner cell spectral representation of the solution, which provides a highly accurate representation and inherently observes (a part of the ) non-local effects of turbulence. These properties have already been successfully exploited in~\cite{laf} to design a filter criterion based on the modal kinetic energy distribution within a cell. In \cite{flad2017} it was shown that the best results are obtained when specifically shaping a plateau-cusp like model dissipation inspired by the spectral eddy viscosity model of Chollet et al.~\cite{chollet1985}. This is however computationally expensive, involving the Variational Multiscale  methodology with its many filter operations (filtering gradients and fluxes, see~\cite{les1} 
for details).

The idea in this paper is to instead design a dissipative filter, which introduces suitable artificial viscosity through the modal distribution of its filter parameters. The filter strength is an additional free parameter of the scheme.
As this procedure is inspired by SPEVM, the filter strength is chosen similar to the scaling of the eddy viscosity, see e.g. \cite{lesieur1996} for an overview, to be
proportional to $(E(k_c,t)/k_c)^{(1/2)}$ as
\begin{equation}
 \sigma_F = c\left(\frac{E(N,t)}{L_{ref}}\right)^{1/2} \frac{1}{\Delta^2},
\end{equation}
with $E(N,t)$ defined as the kinetic energy of the last polynomial mode represented within a cell
\begin{equation}
 E(N,t):= \int\limits_{Q}\tilde{\vec{v}}^2dx \approx \sum\limits_{p,q,r=0}^{N}\;\tilde{\vec{v}}^2J_{pqr}\omega_p^N\omega_q^N\omega_r^N.
\end{equation}

Here $\tilde{\vec{v}}$ is a high pass filtered velocity field obtained by low pass test filtering the velocity within a cell and subtracting the result from 
the unfiltered velocity and Q denotes the volume of a given element. For test filtering, a modal filter is applied according to \eqref{eq:projectionfilter} with filter coefficients $\sigma_{0 \to N-1} 
=1$, $\sigma_N=0$. Test filtering is applied only once per timestep for the sake of computational efficiency (instead of in each RK stage). $J_{pqr}$ denotes the Jacobian of the element transformation to reference space and $\omega^N$ the LGL quadrature weights. $\Delta$ 
is the filter width as used for the Smagorinsky model \cite{flad2017} ($Q/(N+1)^3$) and $L_{ref}=2\pi$ is chosen as a reference length. The 
additional multiplication 
with $1/\Delta^2$ ensures consistent dimensional 
units, while $c$ is a dimensionless constant left for optimization.\\

\subsection{Optimization procedure}
The described sub-grid scale model has $N+1$ free parameters for the filter matrix diagonal entries and additionally the constant $c$. This parameter space is reduced by setting the first entry \red{$\widetilde{K}_{11} = \sigma_0 = 1$} and the last entry \red{$\widetilde{K}_{NN}=\sigma_N=0$}. The rational behind the first choice is that it ensures the conservation of each filtered quantity, as it leaves the first Legendre mode, i.e. the mean per element, unchanged. The last entry of the filter diagonal is set to zero to ensure a cusp-like behavior of the model, as required in the SPEVM. Note that according to the filter described in Sec.~\ref{DGiles}, setting \red{$\widetilde{K}_{ij}=0$} does not eliminate the specific mode, rather this choice selects the highest possible damping for a given filter strength $\sigma_F$. \red{We found that the convergence rate of the scheme remains unchanged, as the filter strength is scaled by the highest modes energy content (decaying fastest as $\Delta x \to 0$), and a full set of modes is always maintained by the procedure.} Finally, the parameter vector to be optimized is $\vec{x}=(c,\sigma_1,...,\sigma_{N-1})$

To find the optimal parameters, data from DNS is utilized. First, a DNS of a DHIT was initialized as described in \red{~\cite{flad2017,flad2018}, with integral length scale $L_{int}\approx1.3$ and $Re_\lambda\approx162$}. The random initial solution is interpolated onto a fine computational grid consisting of $64$ cells per direction and $8$ interpolation LGL points. A DNS is conducted with DGSEM and the solution is afterwards filtered to the LES grid. The filtering procedure consists of a interpolation onto an $8^3$ grid equipped with a sufficiently fine inner cell interpolation grid with LGL nodes, and a subsequent projection to the LES polynomial degree $N=7$. Hence, the LES resolution is $64$ DOF per direction. 
A similar resolution was used in \cite{flad2017} ($48$ DOF). This filtering procedure allows for a direct evaluation of the filtered kinetic energy and other relevant quantities such as the dissipation rate. The objective function is the sum of the $L_2$-errors of the integral kinetic energy and the \red{resolved} integral dissipation rate over 3 test points in time
\begin{equation}
 \vec{x} = arg min \sum_i(E_{kin,\widetilde{DNS}}(t_i) - E_{kin,LES}(t_i))^2 + (\kappa_{\widetilde{DNS}}(t_i) - \kappa_{LES}(t_i))^2,
\end{equation}\red{
for which we compute the resolved dissipation rate of filtered DNS and LES by $\kappa=2\nu\int_{\Omega}S_{ij}S_{ij}d\Omega$, where $S=0.5(\vec{\nabla}\vec{v} + (\vec{\nabla}\vec{v})^T)$ is the strain rate tensor. We note that this is rather a measure for the resolved gradient field than for the kinetic energy decay of the LES as most of the kinetic energy decay stems from model terms.} This function was found to be more effective than using only the kinetic energy. Note that multiple solutions may still exist for the given objective function to be 
minimized.

We optimize the objective function with the Nelder-Mead-method~\cite{nelder1965}. This method is also known as the downhill simplex method, as it spans a simplex consisting of $n+1$ points in $n-$dimensional parameter space. The method was introduced in 1965 by Nelder and Mead and is since used as a simple, heuristic optimization procedure. While there are more efficient optimization procedures, the method has the advantages of being robust and does not require prior knowledge about the objective function or the explicit computation of gradients. Also the requirements to the objective function are low, allowing for discontinuous functions. The method is originally designed for unbounded optimization and is thus augmented by a variable mapping 
\begin{equation}
 \hat{x} = -ln\left(\frac{1}{(\frac{x-b_L}{b_U-b_L})-1}\right),
\end{equation}
transforming the bounded variables $x$ between upper and lower bound $b_U,b_L$, to an unbounded interval, with the inverse mapping given by
\begin{equation}
 x = \frac{b_U-b_L}{1+e^{\hat{x}}} + b_L.
\end{equation}
To avoid getting stuck in local optima, the procedure is reinitialized every $30$ iterations, with a randomly varied simplex centered around the best solution 
at that iteration.\\

\subsection{Optimization results}\label{sec:resultOpt}
\begin{figure}
    \centering
    \includegraphics[width=0.8\textwidth, trim=20 06 60 20 ,clip]{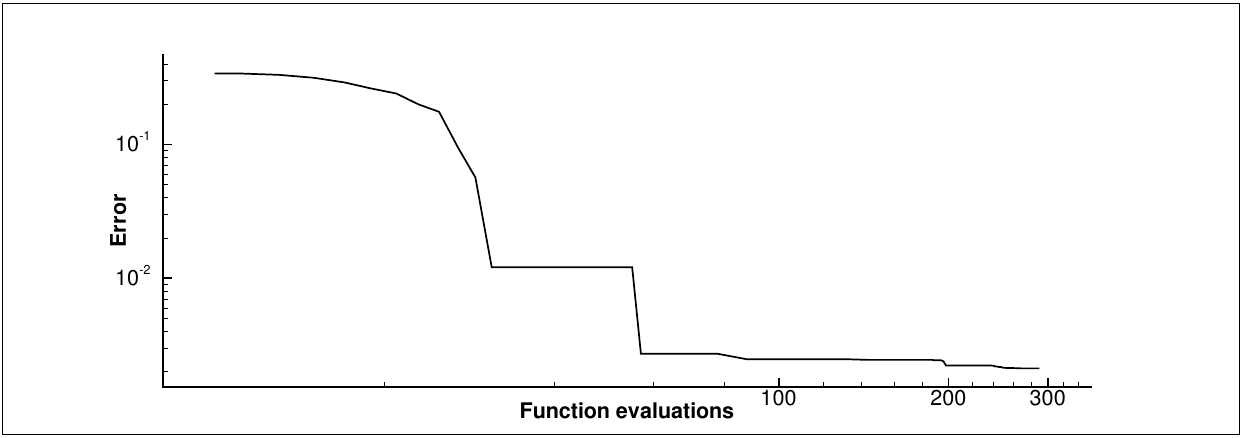}
  \caption{Convergence of the optimization procedure, plotting the error of the objective function over total function evaluations}
  \label{fig:convergence}
\end{figure}

The optimization is run in the regime of exponential decay of turbulence kinetic energy for about $0.5$ large eddy turn over times. On the super computer Cray-XC40 at the HLRS Stuttgart, one run took about 15 seconds on $256$ cores ($2$ cells per core). The optimizer reduced the error by about three orders of magnitude, see Fig.~\ref{fig:convergence}, starting from an initial parameter set $\sigma_{1\rightarrow N-1} = (0.55, 0.55, 0.55, 0.55, 0.55, 0.55, 0.75)$ and $c=1.25$ 
with bounds for all $\sigma_i\in[0.1,1]$ and $c\in[0.1,2]$. The limits of $c$ are chosen motivated by the SPEVM, where the theoretical constant is found for a 
Kolmogorov spectrum to be $\approx0.81$. 

\begin{figure}[!htpb]
    \centering
    \includegraphics[width=0.5\textwidth, height=4.5cm, trim=15 10 60 20 ,clip]{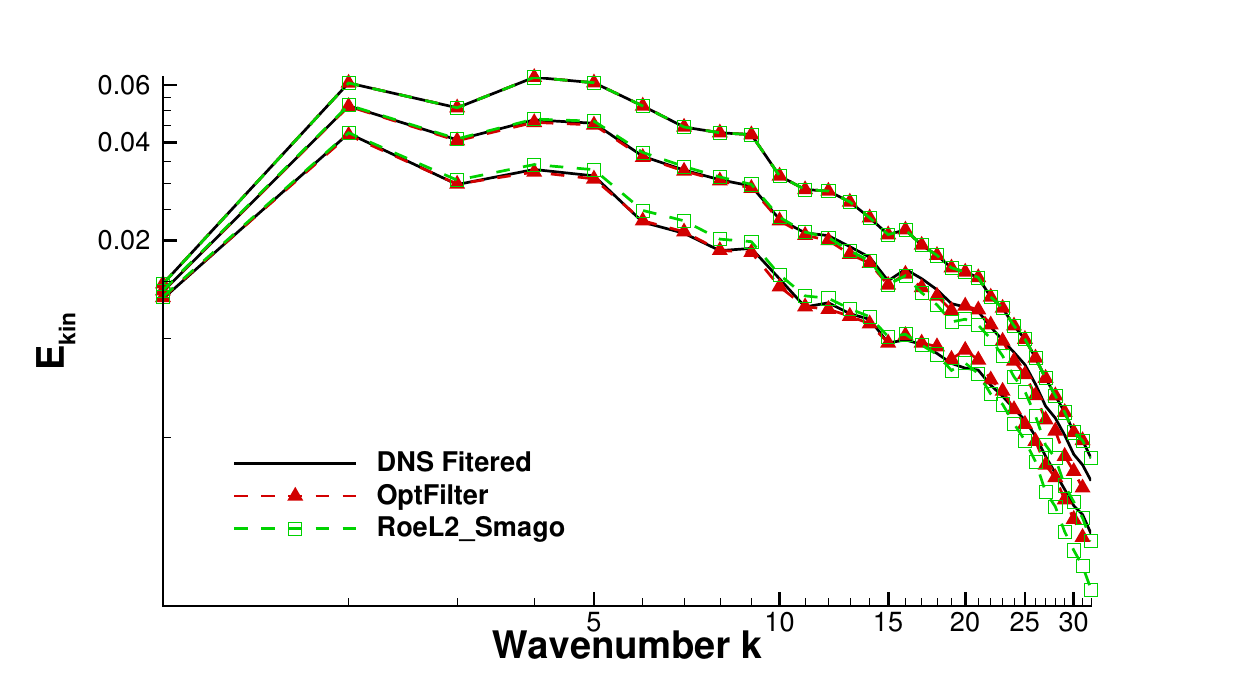}
  \caption{Decaying homogeneous isotropic turbulence filtered DNS/LES kinetic energy spectra: LES with Smagorinsky model and the optimized filter 
dissipation both with $L^2Roe$ Riemann solver. Three different points in time from top to bottom: starting point $t_0$, $t_0+0.2T^*$, $t_0+0.5T^*$}
  \label{fig:spectraOpt}
\end{figure}

The final shape of parameter found by the optimization procedure after $290$ function evaluations is
\begin{align}
 \sigma_{1\rightarrow N-1} = (&0.925, 1.00, 0.853, \\ \nonumber
                              &0.557, 0.889, 0.896)\\ \nonumber
 c = &0.2 \nonumber
\end{align}
Interestingly there is no monotonic decay of the filter matrix coefficients towards higher polynomial modes as could be expected from the SPEVM analogy. As 
described in~\cite{flad2017} the interaction between numerical errors and model dissipation can be complicated, which likely leads to the presented 
results. Fig.~\ref{fig:spectraOpt} shows the kinetic energy spectra resulting from the filter-based approach with optimized parameters in comparison to the 
reference filtered DNS and the best method found in \cite{flad2017}. The filter based approach is found to give at least as good results.

\begin{figure}
    \centering
    \includegraphics[width=0.8\textwidth]{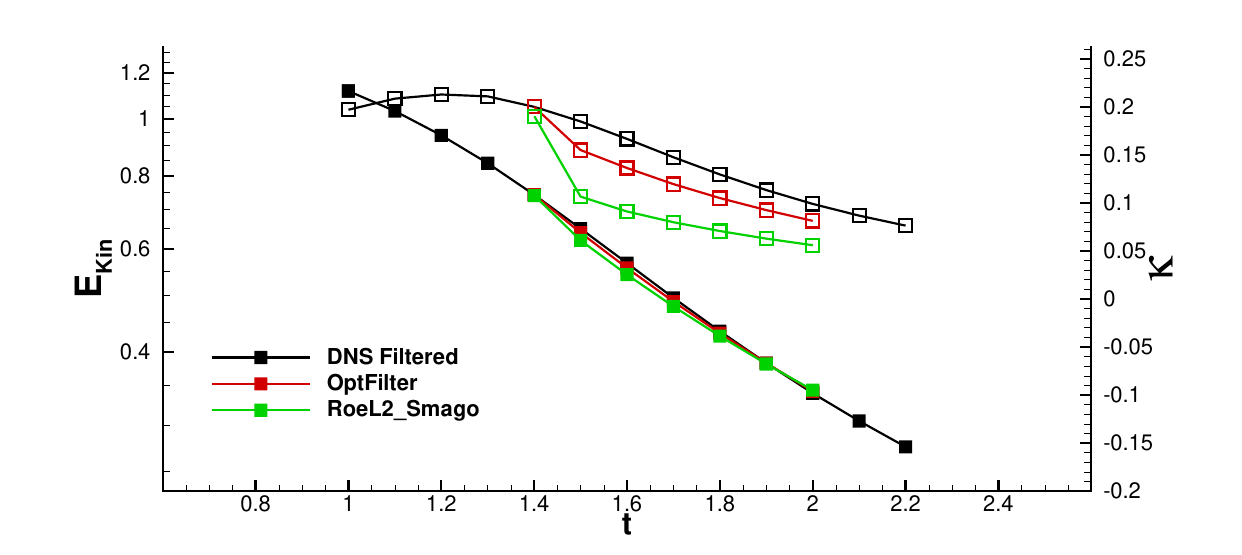}
    \caption{Decaying homogeneous isotropic turbulence filtered DNS/LES, kinetic energy and dissipation over time: LES with Smagorinsky model and the optimized 
             filter dissipation both with $L^2Roe$ Riemann solver. Open symbols are dissipation curves.}
\label{fig:ekinOpt}
\end{figure}

Further analyzing the obtained method, Fig.~\ref{fig:ekinOpt} shows the temporal evolution of the kinetic energy and the dissipation rate $\kappa$. Note that the sum of these two was used as the objective function. It can seen that the kinetic energy of both LES methods follows closely the filtered DNS result, with a small advantage for the filter based method. The dissipation rate of the filter based method is much closer to the reference than the Smagorinsky LES result. 
This shows that the method was able to better resolve the gradients of the flow field.

The described optimization procedure was repeated for polynomial degrees $3$ to $10$. All meshes were chosen such as to obtain approximately $64$ DOF per direction. The resulting filter shapes and strengths are listed in Tbl.~\ref{tab:coeffs}. Note that for the optimization procedure, multiple parameter sets with similar errors may exist due to local minima in the non-convex cost function, hence a clear trend for the filter shapes is not observed.

\begin{figure}[!htpb]
    \centering
    \includegraphics[width=0.49\textwidth, height=4.5cm, trim=15 10 60 20 ,clip]{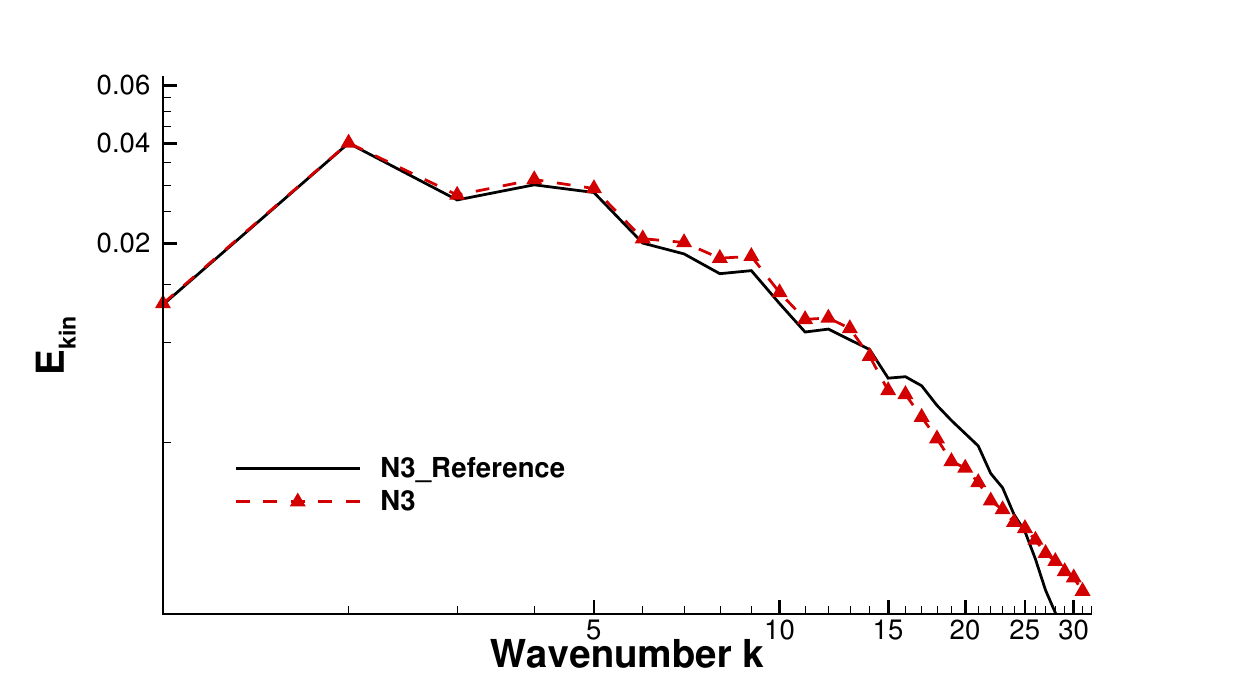}
    \includegraphics[width=0.49\textwidth, height=4.5cm, trim=15 10 60 20 ,clip]{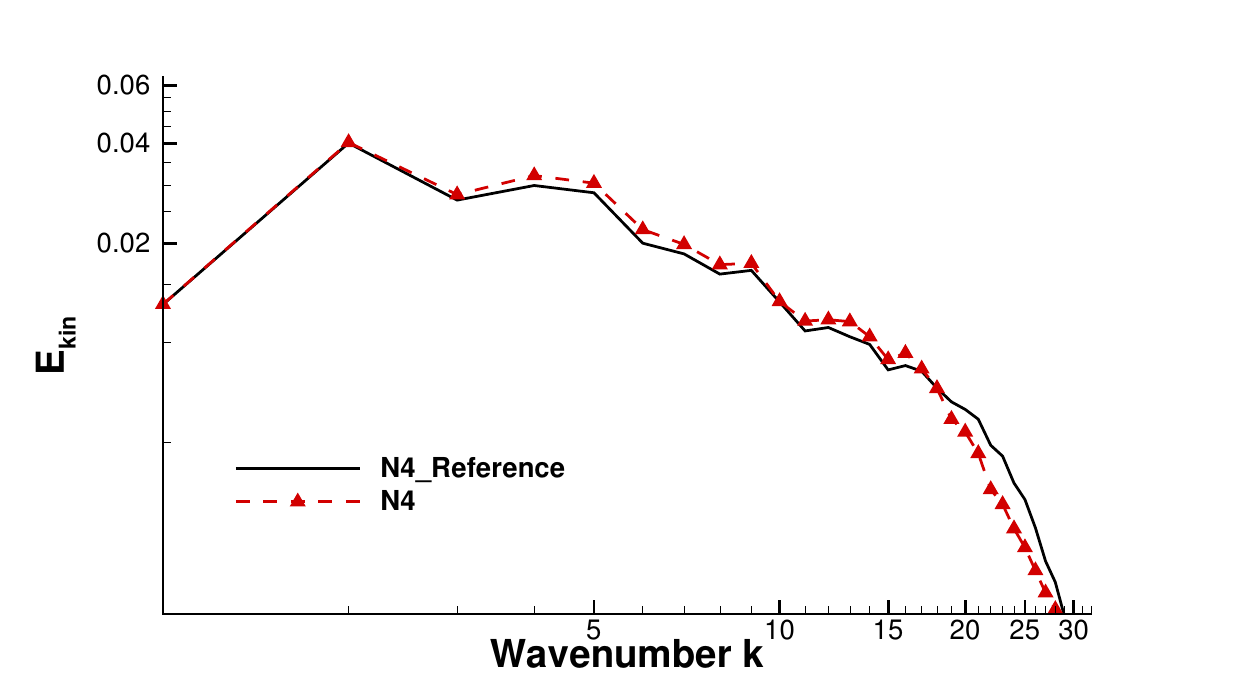}\\
    \includegraphics[width=0.49\textwidth, height=4.5cm, trim=15 10 60 20 ,clip]{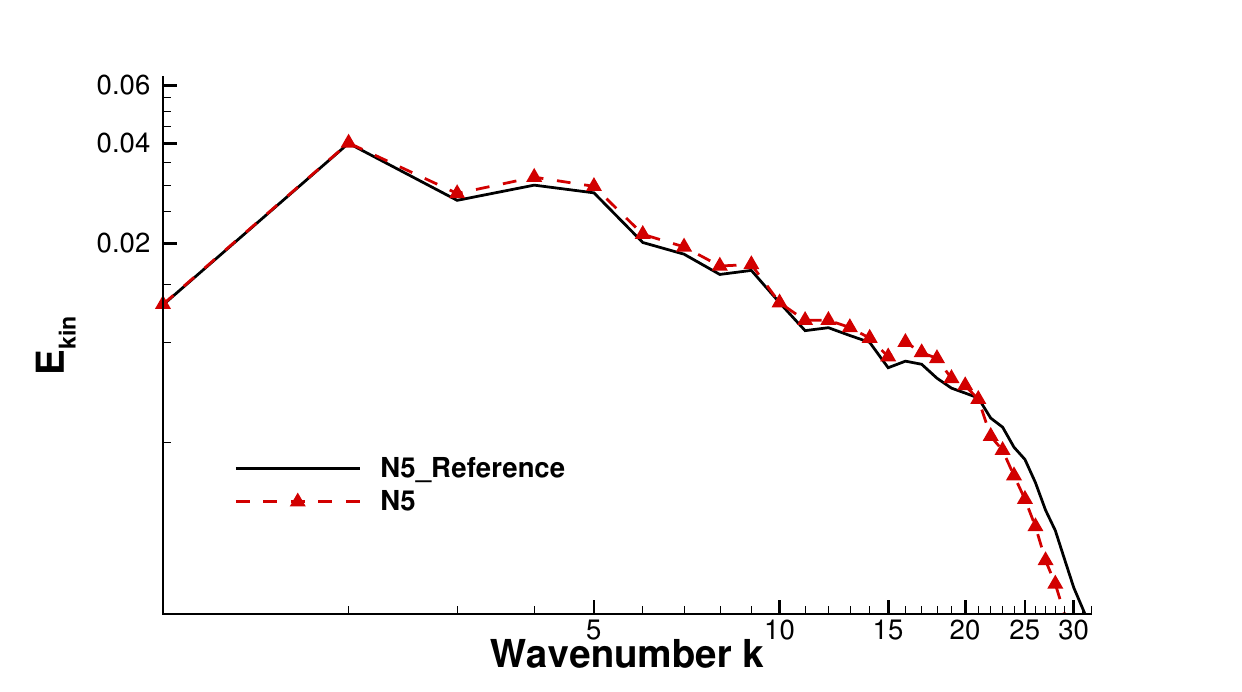}
    \includegraphics[width=0.49\textwidth, height=4.5cm, trim=15 10 60 20 ,clip]{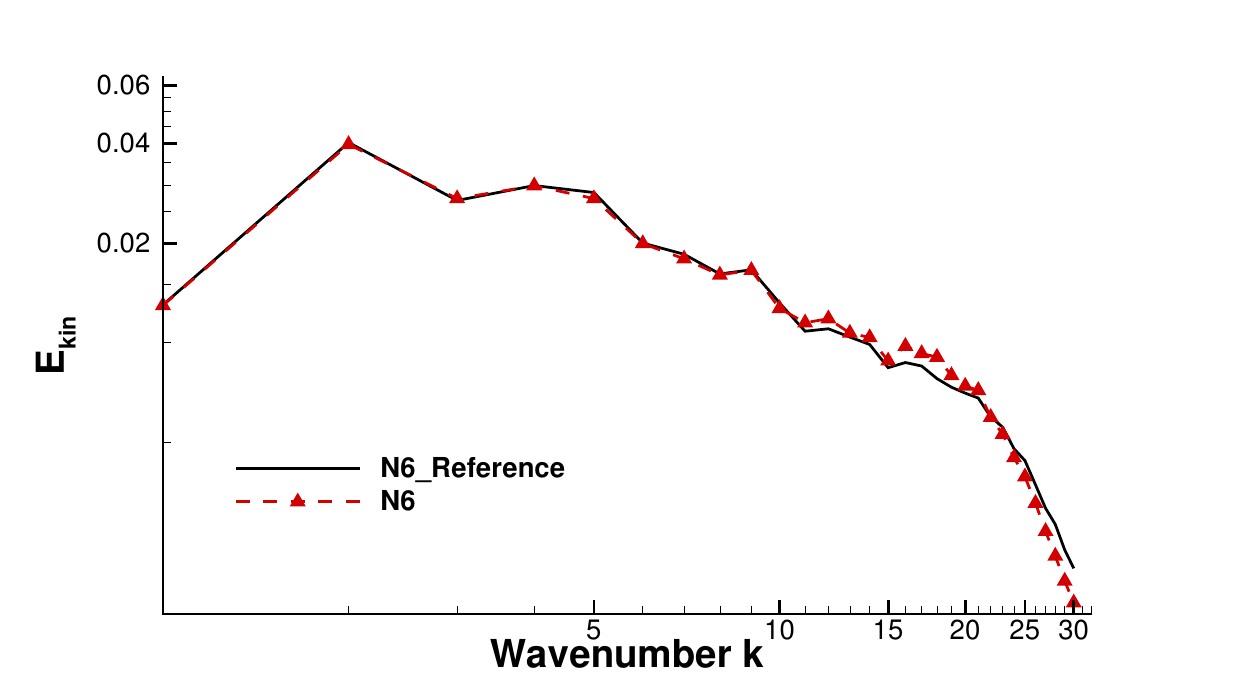}\\
    \includegraphics[width=0.49\textwidth, height=4.5cm, trim=15 10 60 20 ,clip]{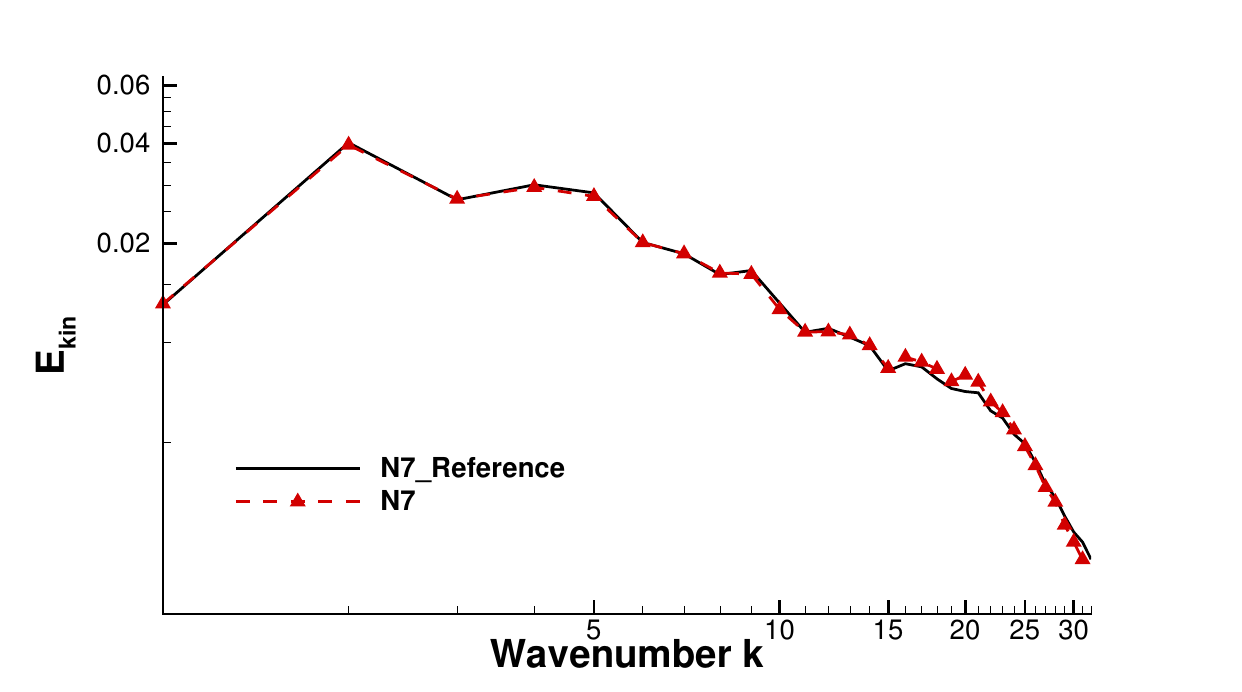}
    \includegraphics[width=0.49\textwidth, height=4.5cm, trim=15 10 60 20 ,clip]{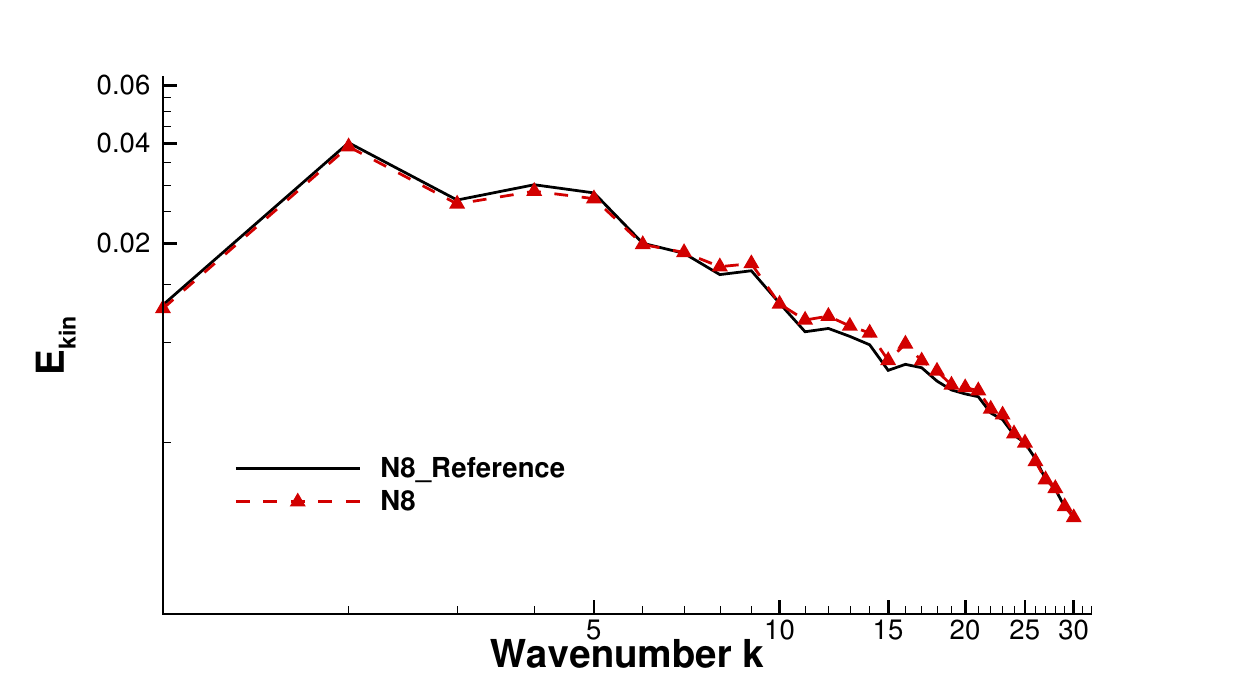}\\
    \includegraphics[width=0.49\textwidth, height=4.5cm, trim=15 10 60 20 ,clip]{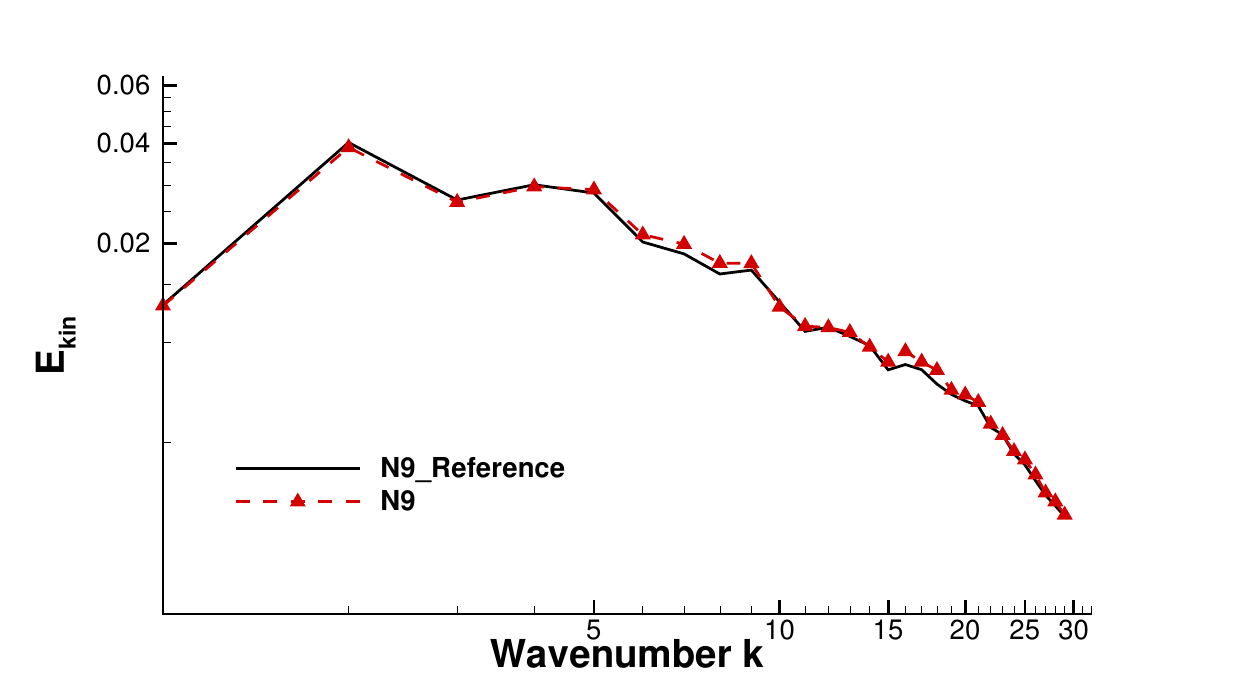}
    \includegraphics[width=0.49\textwidth, height=4.5cm, trim=15 10 60 20 ,clip]{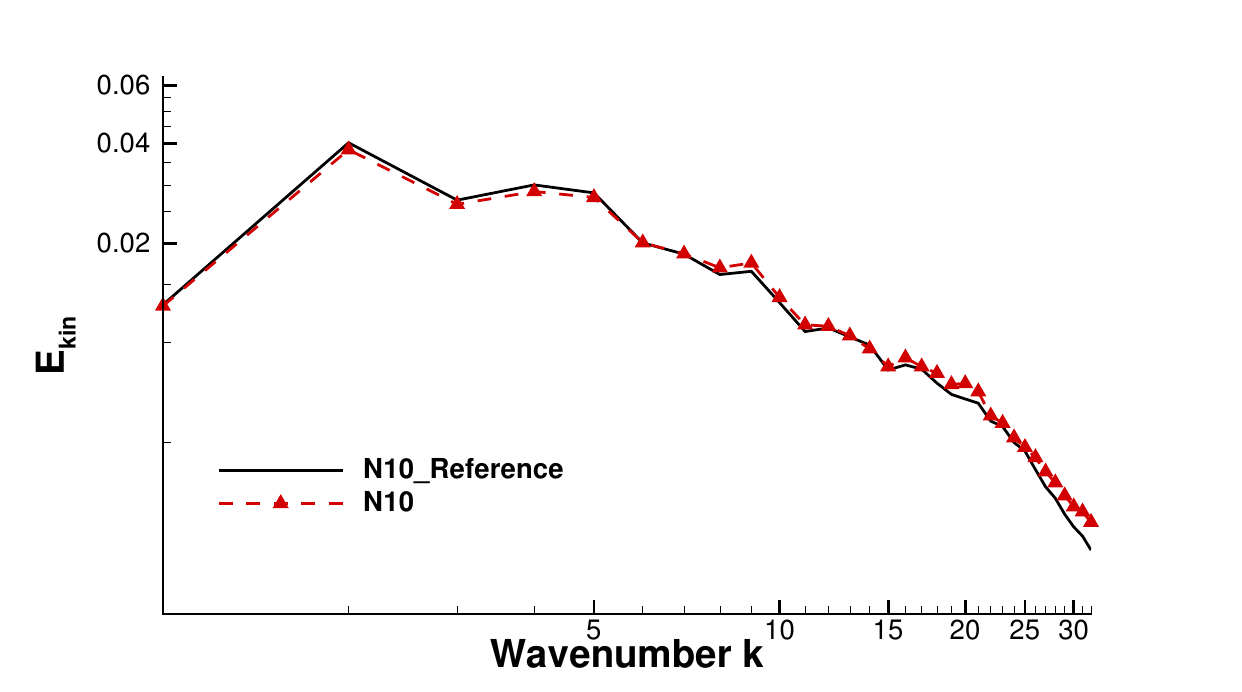}
  \caption{Decaying homogeneous isotropic turbulence filtered DNS/LES kinetic energy spectra, $t_0+0.6T^*$}
  \label{fig:spec_all}
\end{figure}

\begin{table}
 
\centering{
\tiny{
\begin{tabular}{ccccccccccccccc}
\hline
 = & $\sigma_0$ & $\sigma_1$ & $\sigma_2$ & $\sigma_3$ & $\sigma_4$ & $\sigma_5$ & $\sigma_6$ & $\sigma_7$ & $\sigma_8$ & $\sigma_9$ & $\sigma_{10}$ & $c$ & 
$c_{\infty}$&$\mathrm{Error}[E-2]$\\
 \hline
 N3   & 1 & 0.799 & 0.656 & 0 & - & - & - & - & - & - & -& 0.061 & 0.061&4.11\\
N4    & 1 & 1.00 & 0.01 & 1.00 & 0 & - & - & - & - & - & -& 0.11 & 0.11&4.21\\
N5    & 1 & 1.00 & 0.623 & 0.991 & 1.00 & 0 & - & - & - & - & -& 0.202 & 0.2&4.5\\
N6    & 1 & 0.873 & 0.846 & 1.00 & 0.304 & 0.07 & 0 & - & - & - & -& 0.132&0.31&3.27\\
N7    & 1 & 0.925 & 1.00 & 0.853 & 0.557 & 0.889 & 0.896 & 0 & - & - & -& 0.2&0.35&2.76\\
N8    & 1 & 0.939 & 0.973 & 1.00 & 0.915 & 0.903 & 0.157 & 0.985 & 0 & - & -& 0.237&0.35&3.22\\
N9    & 1 & 0.958 & 1.00 & 1.00 & 0.629 & 0.832 & 1.00 & 1.00 & 0.01 & 0 & -& 0.250&0.35&2.93\\
N10   & 1 & 0.957 & 0.989 & 0.999 & 1.00 & 0.632 & 0.838 & 1.00 & 1.00 & 0.01 & 0& 0.25&0.38&2.49\\
 \hline

 \end{tabular}
}}
 \caption{Results of the optimization procedure: filter coefficients $\sigma_i$, $c$ and $c_{\infty}$ is the 
filter strengths for DHIT and infinite Reynolds number test case (see Sec.~\ref{sec:infty}) respectively and $\mathrm{Error}$ is the final value of the objective 
function after 
optimization.}
\label{tab:coeffs}
\end{table}
For all polynomial degrees the final spectra resulting from a computation with the optimized filter are shown in Fig.~\ref{fig:spec_all} in comparison to the 
respective filtered DNS on the same grid. All 
results give excellent spectra and can thus be considered a good outcome of the optimization procedure. It is found that the 
optimization works slightly better for higher polynomial degrees. This can be expected as the parameter space for the optimization is larger. 

\section{Testing for $Re\rightarrow\infty$}\label{sec:infty}
\subsection{Decaying HIT}
\begin{figure}
  \subfigure[]{%
    \includegraphics[width=0.5\textwidth, height=4.5cm, trim=20 10 60 20 ,clip]{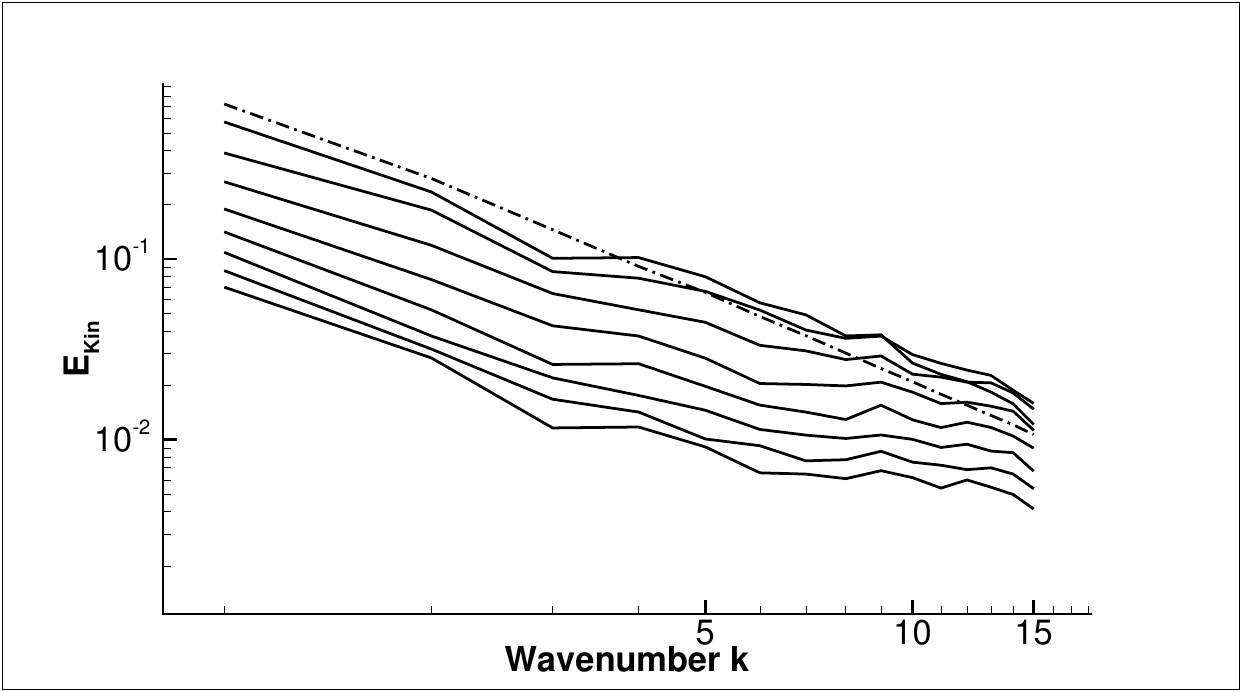}}
  \subfigure[]{%
    \includegraphics[width=0.5\textwidth, height=4.5cm, trim=20 10 60 20 ,clip]{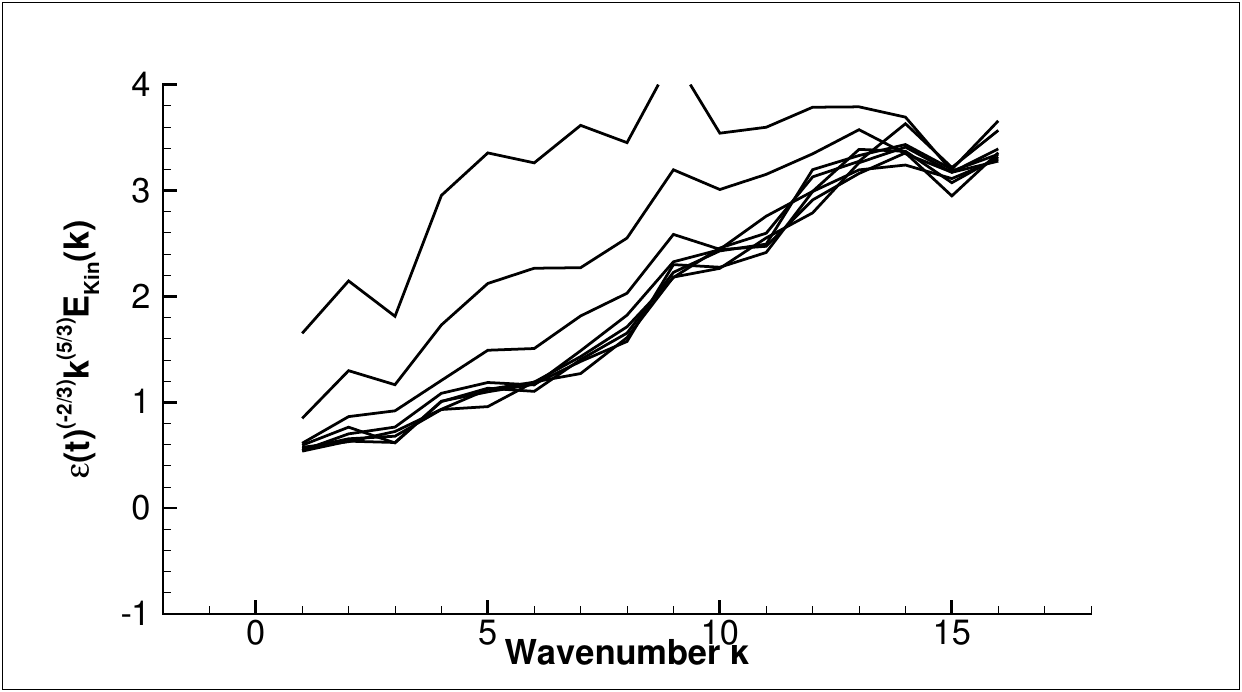}}\\
  \caption{Left: kinetic energy spectra at different points in time. Right: Kolmogorov constant. No-model LES using Roe's Riemann solver}
  \label{fig:spec_roe}
\end{figure}

As reported in \cite{flad2017}, for severely underresolved flow fields it is insufficient to use only the Riemann solver dissipation as a model 
surrogate. In contrast using KEP-fluxes with Smagorinsky's model worked reasonably well. In the preceding section a model based on an 
optimized 
dissipative filter shape for LES was introduced. In this section, the described approaches will be tested against decaying HIT in the limit of vanishing 
viscosity. The 
simulation is initialized as described in~\cite{flad2017}, but with a constant $k^{-5/3}$ distribution of kinetic energy up to $k_c=16$. The test setup is 
simulated with $6$ cells per direction and polynomial degree $N=7$, leading to a grid Nyquist wavenumber of $24$, and a resolution limit considering $3$ points 
per wavelength of $k_c=16$. The test case is particularly interesting as it constitutes the highest degree of under-resolution possible. Also, if a model is found that 
needs no computation of the gradients for this test case, gradient computation can be skipped altogether, leading to a much cheaper numerical method (about 
$50\%$ of DGSEM is used for second order terms). The filter-based model has that property, not needing any computation of gradients for introducing the 
artificial dissipation.\\
In contrast to the DHIT test case in Sec.~\ref{sec:resultOpt}, the exact solution cannot be computed by means of DNS as the required resolution goes 
towards infinity. Instead, the results are compared to theoretical findings. First the kinetic energy spectra are examined. The kinetic energy in this test 
case 
is considered to decay self similarly, i.e. the spectra should all have a $k^{-5/3}$ slope as is the case for the initial solution. Secondly, the compensated 
kinetic energy spectra normalized by $\epsilon(t)^{(-2/3)}k^{(5/3)}$ (here \red{$\epsilon(t)=-dE_{kin}(t)/dt$}) should collapse to a constant in an ideal setup, known as the Kolmogorov constant. Values for the Kolmogorov constant vary in literature but are usually found to be around $1.6$, an overview is found in ~\cite{Yeung97}. The normalized spectra have the 
advantage that the decay rate in time and the spectral shape of kinetic energy can be analyzed in one function. Fig.~\ref{fig:spec_roe} shows the 
results for the no-model LES using Roe's Riemann solver, confirming the previous findings of its inability to give accurate results for high levels of 
under-resolution~\cite{flad2017}. As already observable through the upward bend of the spectra, the Kolmogorov normalized spectra clearly show that the method leads to non-physical behavior, failing to 
obtain a plateau.  

\begin{figure}
  \subfigure[]{%
    \includegraphics[width=0.5\textwidth, height=4.5cm, trim=20 10 60 20 ,clip]{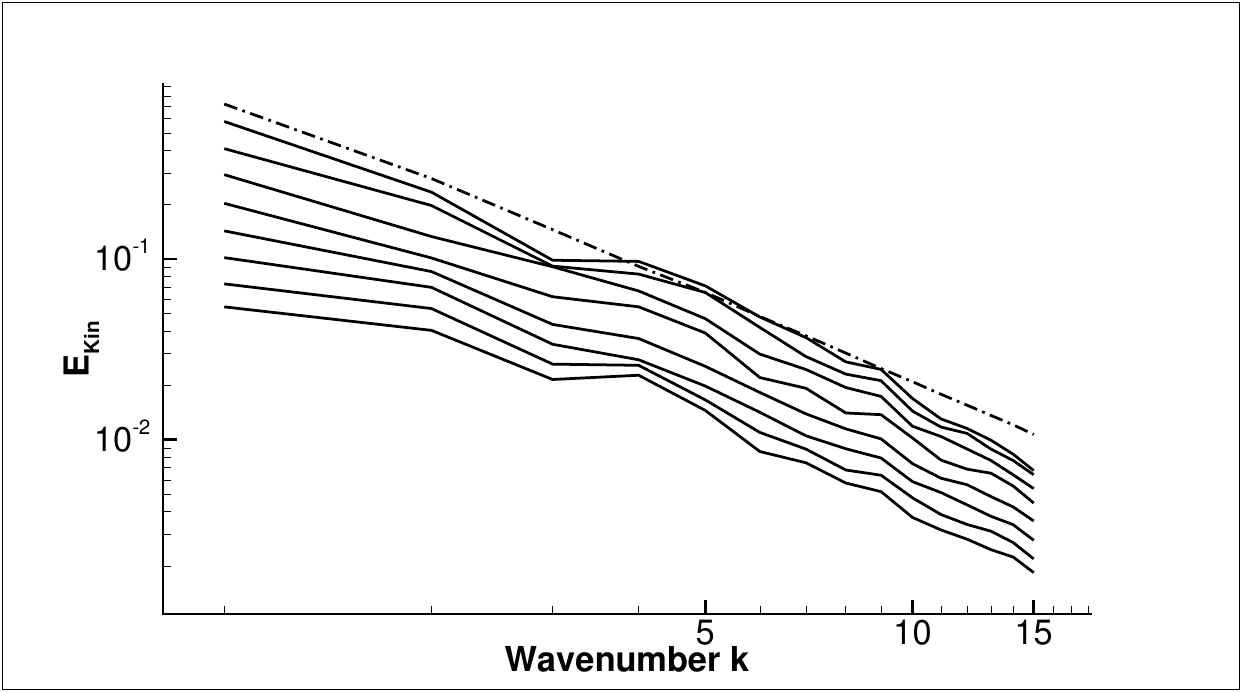}}
  \subfigure[]{%
    \includegraphics[width=0.5\textwidth, height=4.5cm, trim=20 10 60 20 ,clip]{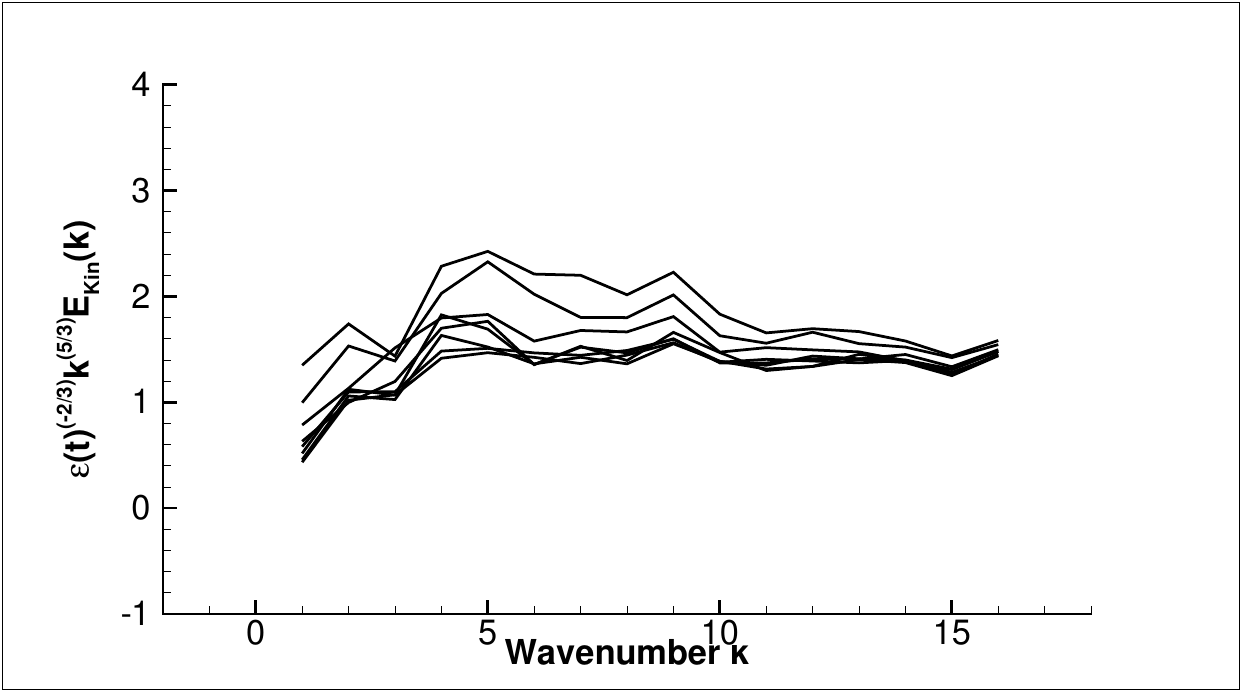}}\\
  \caption{Left: kinetic energy spectra at different points in time. Right: Kolmogorov constant. LES using Smagorinsky's model ($C_S=0.15$) and 
$L^2Roe$ Riemann solver}
  \label{fig:spec_smago}
\end{figure}

Fig.~\ref{fig:spec_smago} shows the results for the superior method found in \cite{flad2017}, but with a Smagorinsky constant 
adjusted to $0.15$ as to obtain optimal results for the selected test case. The need for a case by case adjustment of the Smagorinsky constant is a well known drawback of the 
model, which can be circumvented by using the dynamic model variant. Fig.~\ref{fig:spec_smago} confirms the results also in the limit of infinite Reynolds number. After an initial phase, where 
the random field adjusts to NS 
dynamics, the spectra decay self similarly, maintaining a $k^{-5/3}$ slope up to the highest wavenumbers. The Kolmogorov function exhibits a plateau around 
$1.4$, 
in good accordance with theoretical predictions. 

\begin{figure}
  \subfigure[]{%
    \includegraphics[width=0.5\textwidth, height=4.5cm, trim=20 10 60 20 ,clip]{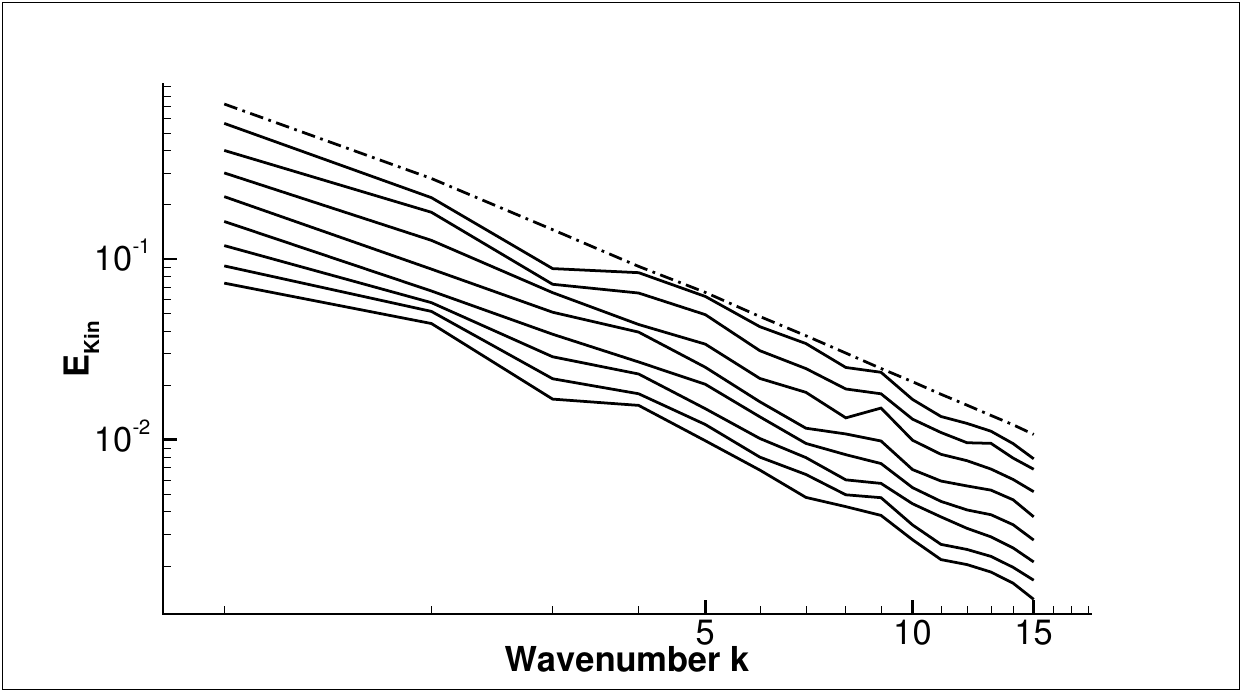}}
   \subfigure[]{%
    \includegraphics[width=0.5\textwidth, height=4.5cm, trim=20 10 60 20 ,clip]{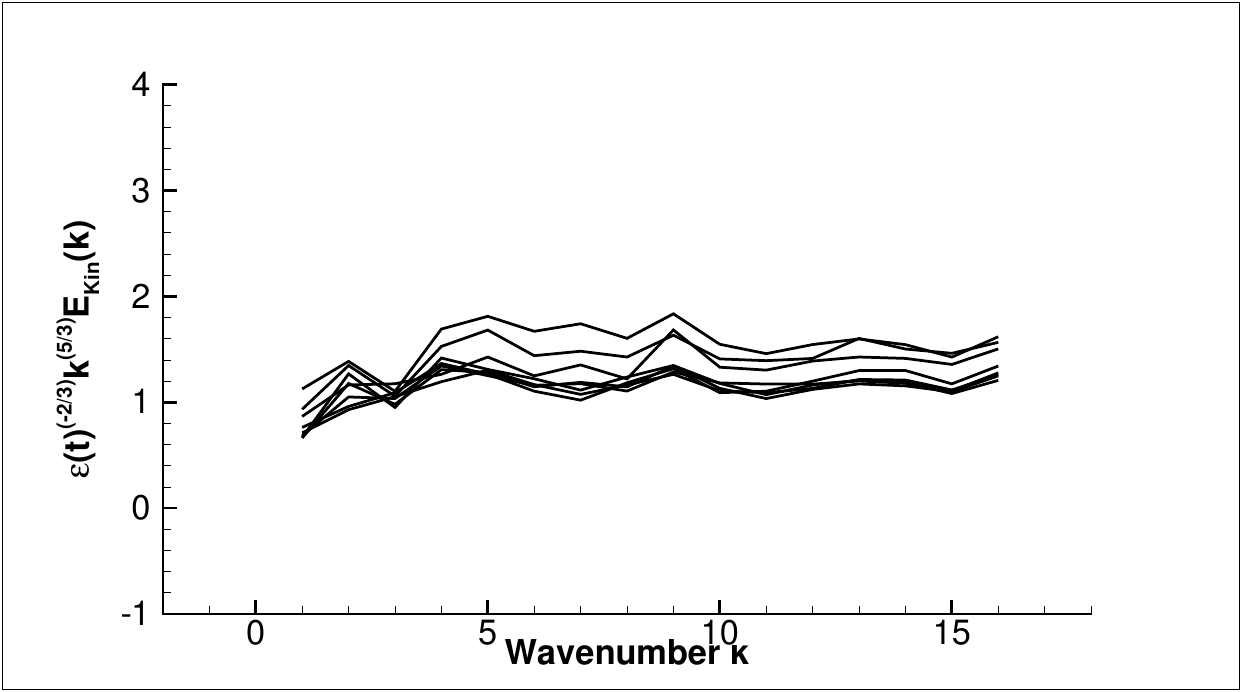}}\\
  \caption{Left: kinetic energy spectra at different points in time. Right: Kolmogorov constant. LES using the described filter based model.}
  \label{fig:spec_fiopt}
\end{figure}

The results for the filter-based LES found in this section are shown in Fig.~\ref{fig:spec_fiopt}. Similar to the Smagorinsky model, the 
constant 
of the filter strength had to be adjusted for the best results to $0.35$. Assuming that the constant is universal, this means that the model reference length 
for this flow is smaller by a factor of $0.566$. The reference length withing the optimization was chosen somewhat arbitrarily. Based on the integral length 
scale of the flow with a ratio of about $\pi \backslash (2\pi) = 0.5$ the change of the constant can be further motivated. The spectra show the best parallel, self similar 
decay of the three methods under 
investigation. The Kolmogorov function plateaus at around $1.2$, which is still in agreement with values reported in literature.

In summary, for decaying homogeneous isotropic turbulence at infinite Reynolds number, the approach of no-model LES is not suitable. Smagorinsky's model and 
the 
filter based approach both with the low Mach $L^2Roe$ Riemann solver give excellent results.

\subsection{Inviscid \blue{Taylor-}Green Vortex}
\begin{figure}
  \subfigure[]{%
    \includegraphics[width=0.5\textwidth, height=4.5cm, trim=20 10 60 20 ,clip]{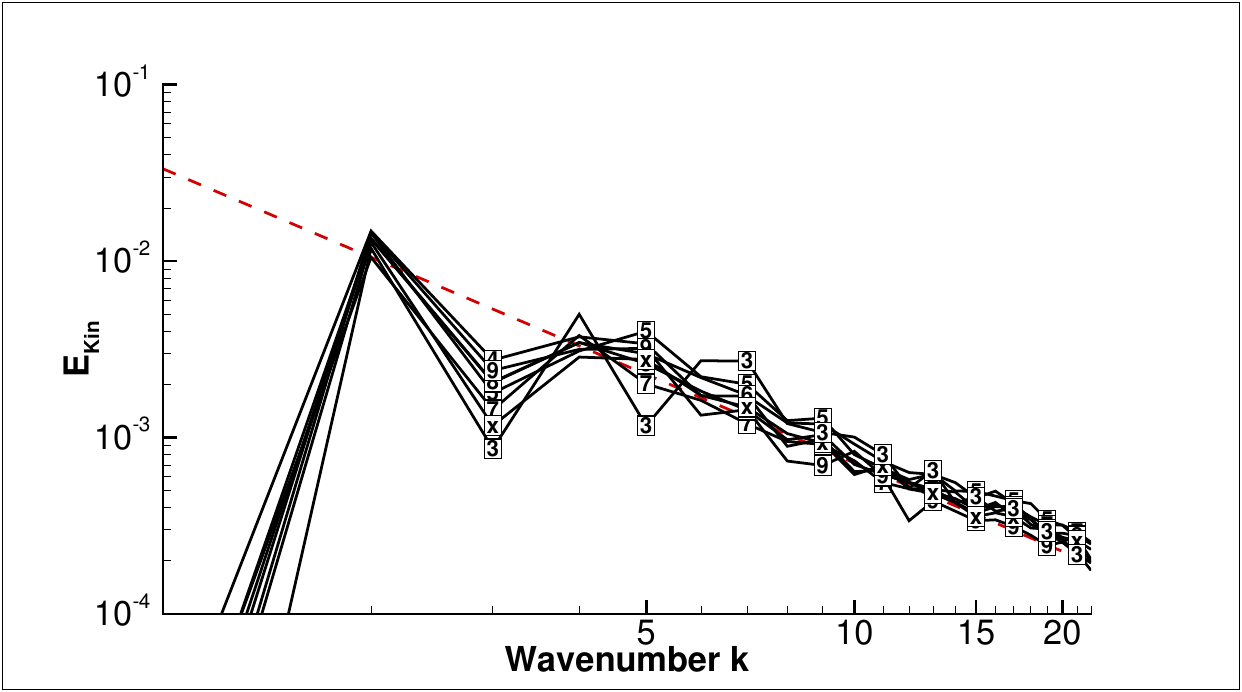}}
   \subfigure[]{%
    \includegraphics[width=0.5\textwidth, height=4.5cm, trim=20 10 60 20 ,clip]{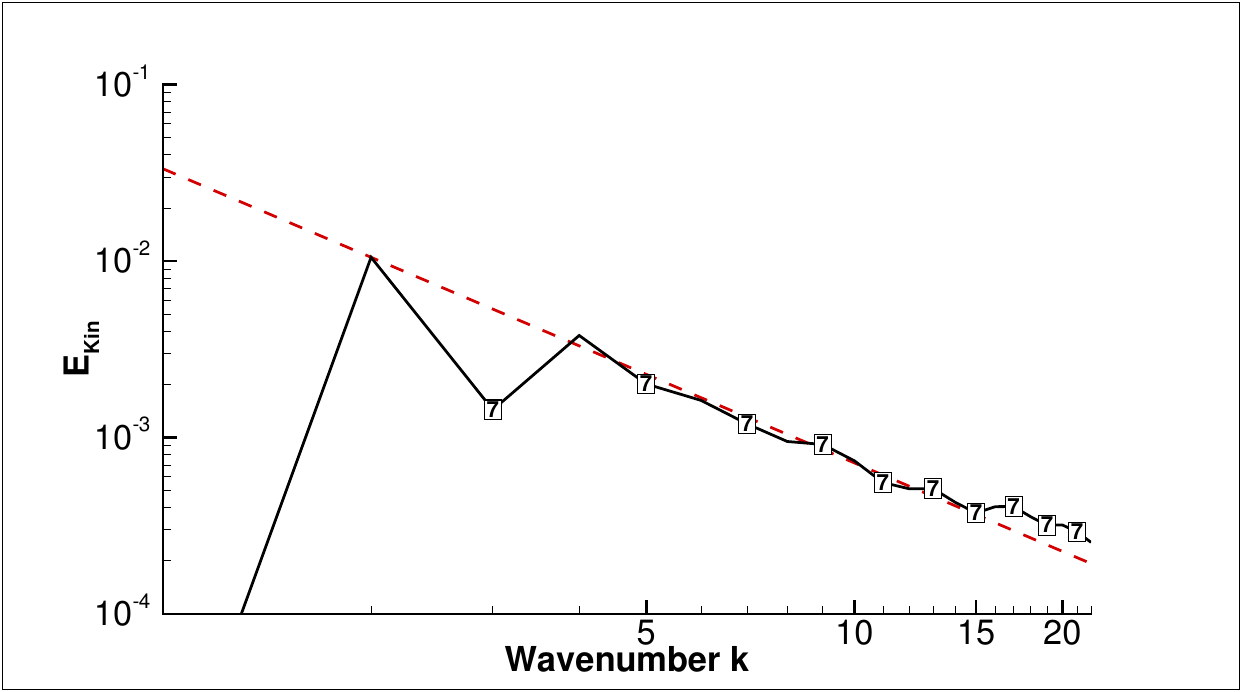}}\\
  \caption{Left: kinetic energy spectra at $t=14$ for different polynomial degrees denoted by symbols. Right: as left but only $N=7$ shown.}
  \label{fig:tgv_spec}
\end{figure}

The inviscid Taylor Green Vortex flow (TGV, see e.g. \cite{Fauconnier2008},\cite{Brachet1991},\cite{Hickel2006},\cite{Brachet1983}) test case is 
chosen to show the behavior
of
the simulations for flows that involve strong turbulent production and transition
mechanisms. The flow is calculated forward in time from its ``laminar'' initial state, up to the point where dissipation peaks and scale separation 
is maximum followed by a subsequent self-similar decay ideally maintaining a $k^{-5/3}$ slope in the kinetic energy spectra. For this inviscid case, the peak and decay are the direct result of the 
numerical (model) viscosity of the method. Thus, the behavior induced by the model in this regime is a suitable metric of model quality. The test case is also easy to set up and compute and is therefore widely used.\\
In this work it is used to show the ability of some LES methods investigated to not introduce dissipation for well resolved flows, 
specifically 
during the initial transition phase. As such it was also used recently in \cite{manzanero2018}. The decay of turbulence after the dissipation peak is 
investigated by means of kinetic energy spectra and normalized spectra as defined in the previous section. Fig.~\ref{fig:tgv_spec} shows the resulting spectra 
(up to $3$ PPW)
at $t=14$ for all polynomial degrees used for optimization (left) and separately for $N=7$ (right), note that the constant $c$ used is the one for infinite 
Reynolds number as denoted in Tbl.\ref{tab:coeffs}. It is found that for all polynomial degrees the filter based LES is able to nicely maintain a $k^{-5/3}$ 
slope in the spectra.  

\begin{figure}
  \subfigure[]{%
    \includegraphics[width=0.5\textwidth, height=4.5cm, trim=20 10 60 20 ,clip]{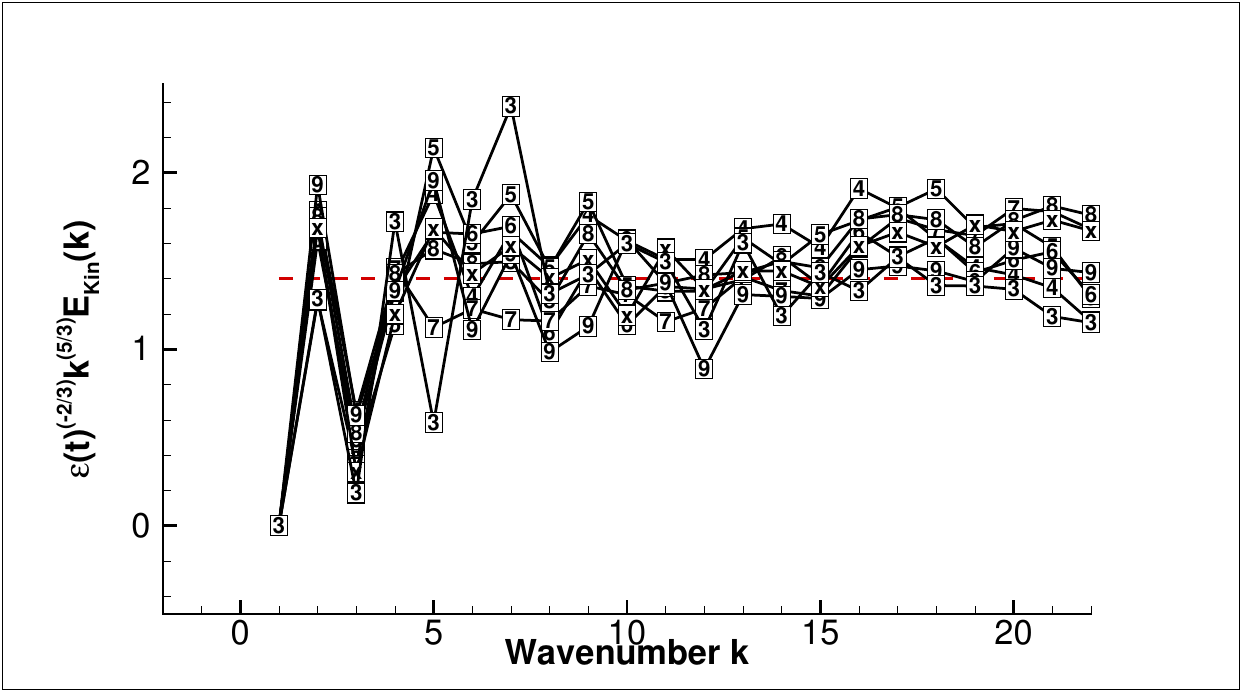}}
   \subfigure[]{%
    \includegraphics[width=0.5\textwidth, height=4.5cm, trim=20 10 60 20 ,clip]{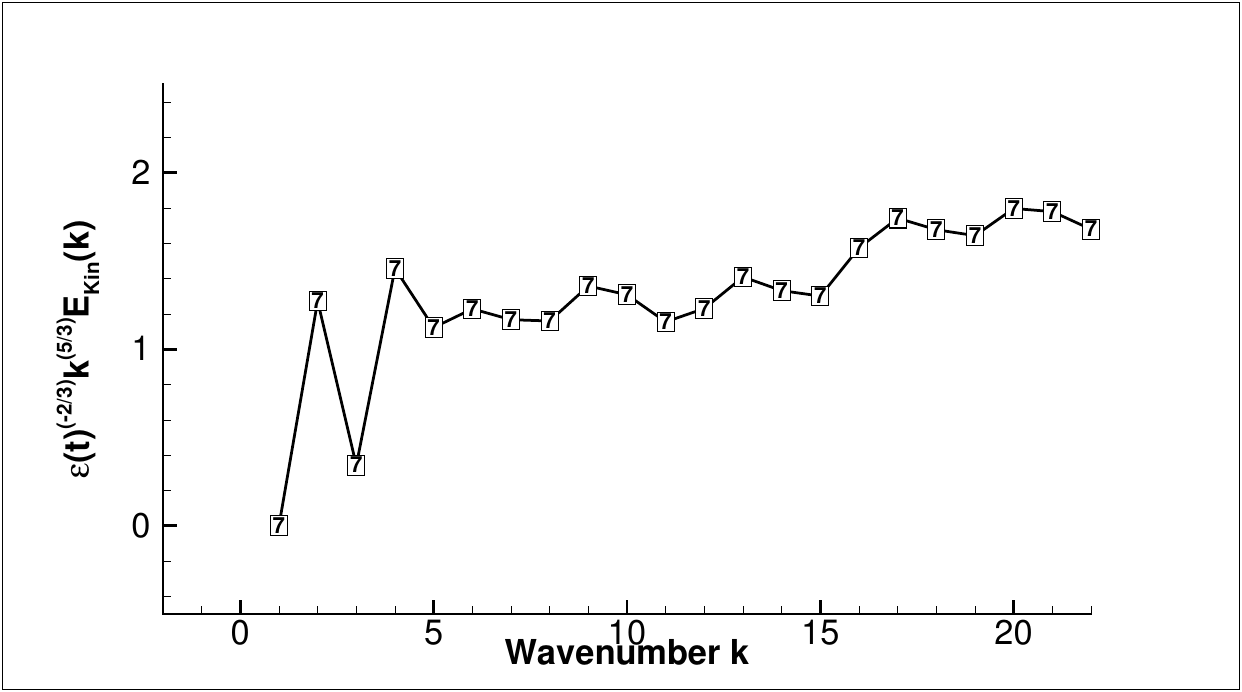}}\\
  \caption{Left: normalized kinetic energy spectra at $t=14$ for different polynomial degrees denoted by symbols. Right: as left but only $N=7$ shown.}
  \label{fig:tgv_kolmo}
\end{figure}

\begin{figure}
    \includegraphics[width=0.5\textwidth, height=4.5cm, trim=20 10 60 20 ,clip]{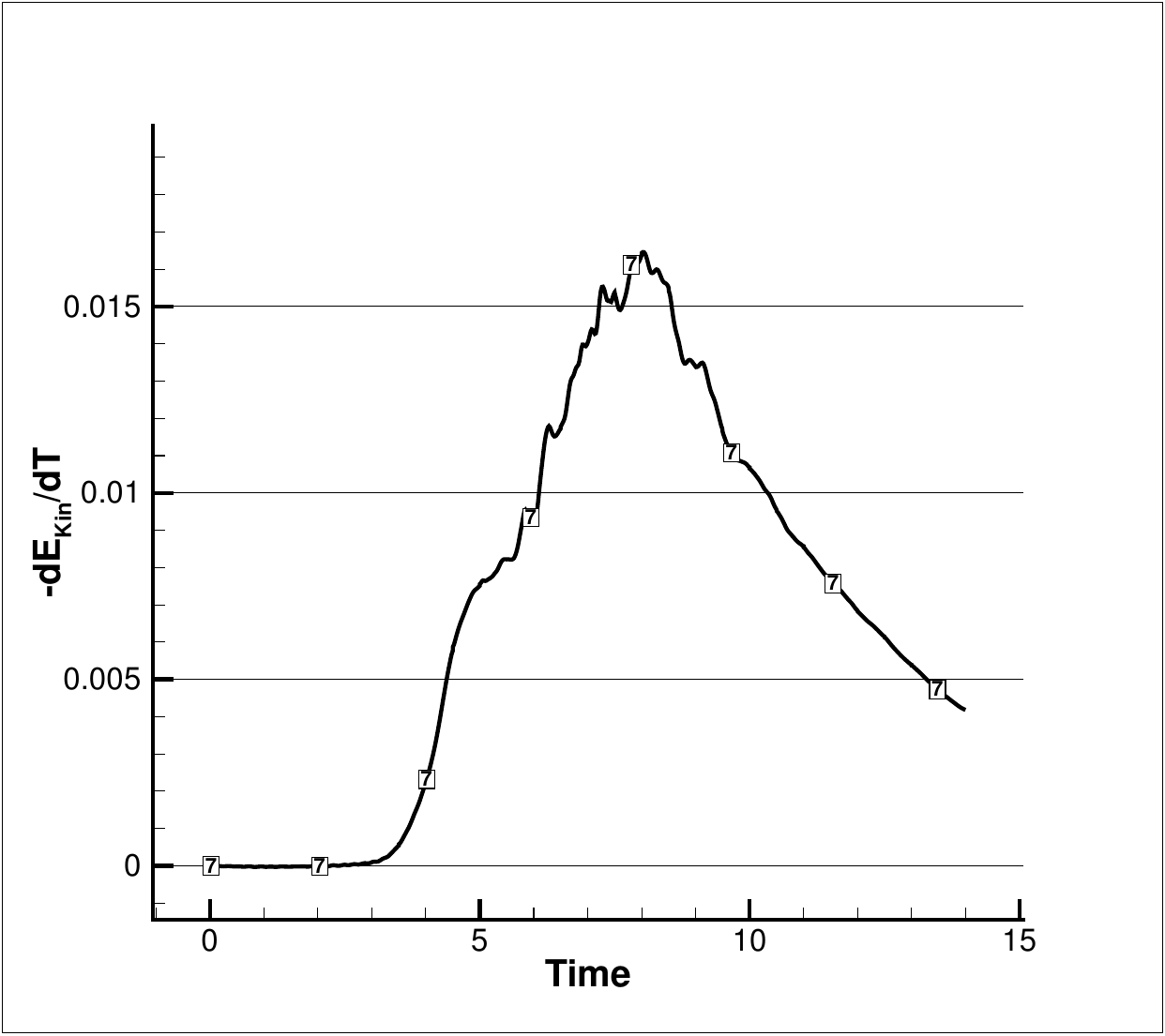}
  \caption{Dissipation rate over time for $N=7$, computed by differentiating integral kinetic energy over time}
  \label{fig:tgv_diss}
\end{figure}

More information than in the usually shown kinetic energy spectra is contained in normalized spectra, as discussed above. This is especially important for the TGV 
testcase, as for infinite Reynolds number no DNS reference can be provided. Fig~\ref{fig:tgv_kolmo} shows the normalized spectra for all polynomial degrees 
(right) and again for $N=7$ separately (left). All normalized spectra show a plateau over a wide range of wavenumbers. The Kolmogorov constant is approximate 
$1.4$ in excellent accordance with theoretical predictions.\\
Finally the dissipation rate calculated as $-dE_{kin}/dt$ is shown in Fig~\ref{fig:tgv_diss}, exemplary for $N=7$ while all other polynomial degrees show 
similar behavior. It is found that the filter procedure does not add considerable damping at the beginning of the computation where the flow field is 
essentially laminar and hence no model viscosity should be added. The dissipation rate shows the typical rapid onset of dissipation once the turbulence 
production started ($t\approx3$). It can thus be 
concluded that the filter based LES allows for clean turbulent transition, while maintaining the expected behavior of self similar decay. The reason it does so is that the filter strength is scaled by the kinetic energy 
content of the last polynomial mode. This energy content decays spectrally whenever the flow is well resolved.

\section{Channel flow at $Re_\tau=590$}\label{sec:channel590}
\begin{figure}
\centering
    \includegraphics[width=0.8\textwidth, trim=20 10 50 20 ,clip]{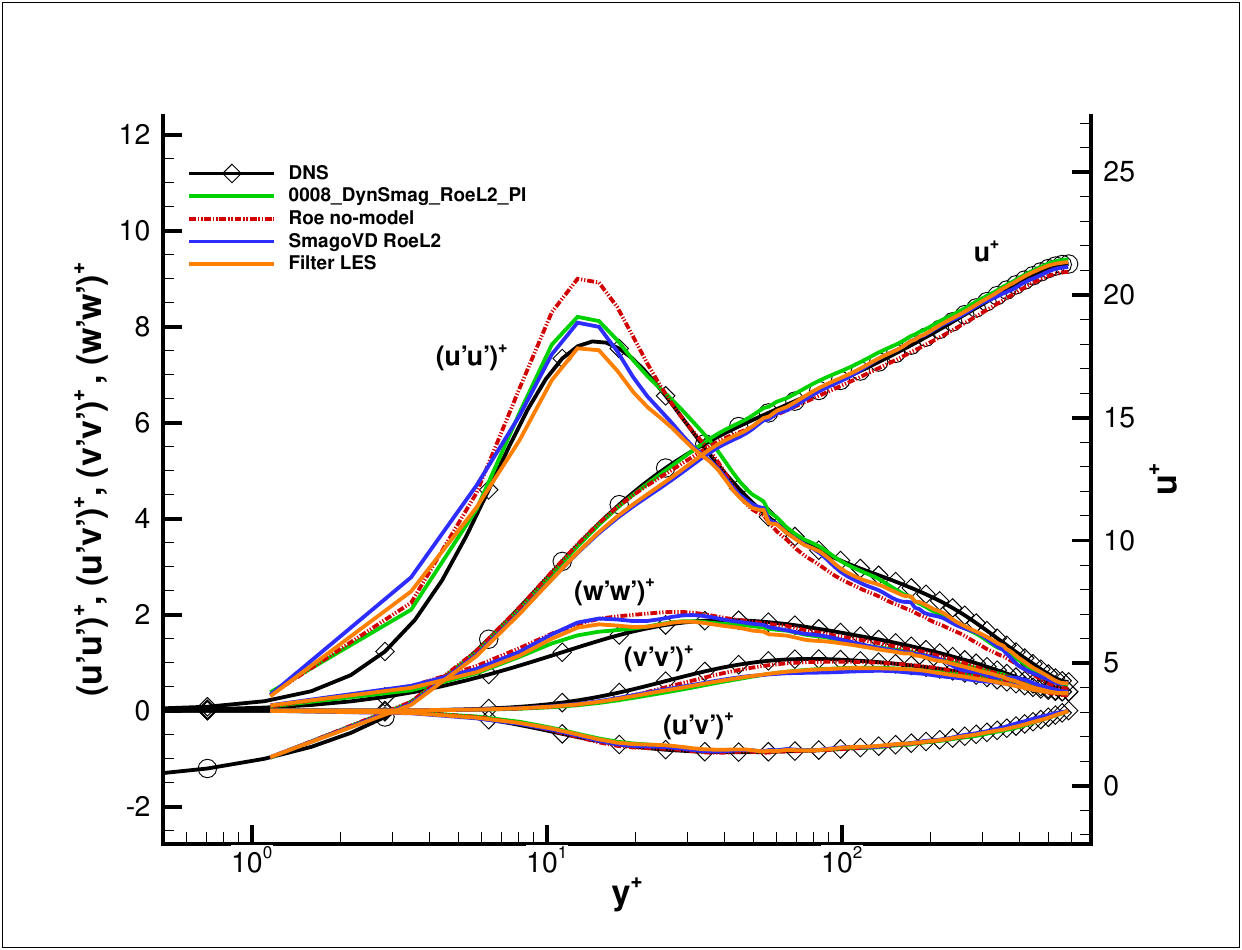}  \caption{Plane turbulent channel flow, 
$Re_{\tau}=590$, $8$ cells per direction, $N=7$. Green: splitDG with dynamic Smagorinsky model. Red: standard implicit LES polynomial de-aliasing 
and Roe's Riemann solver. Blue: splitDG with Smagorinsky and Van-Driest damping. Orange: filter based LES. Reference DNS is taken from Moser 
\cite{moser1999}}
\label{fig:channel}
\end{figure}

So far all numerical experiments focused on DHIT or TGV, both with periodic boundary conditions. For the evaluation of LES model performance, these are the test cases of choice as they offer insights into the 
dissipation behavior of the model. To analyze wall-bounded flows with strong an isotropy of turbulence stresses at the wall, plane turbulent channel flow provides 
a simple setup. For this case, there is usually one direction, the wall normal one, which is significantly better resolved 
than the other two. For wall resolved LES that is necessary to resolve the wall gradient. However, that is by design a contrast to the usual requirements of 
LES models, which require the resolution of energy containing scales only. Most LES models therefore use an ad-hoc adjustment to turn off the model effect nearing the wall. 
Nevertheless most engineering flows are wall 
bounded, and it is thus necessary to check the model behavior for such cases. In this section the three models of the previous section are tested for a wall-resolved LES against a 
plane turbulent channel flow at friction Reynolds number $Re_{\tau}=590$. The no-model LES needs no adjustment, as the increased wall-normal resolution reduces the discretization influence. The Smagorinsky model is used in its dynamic 
variant as described in \cite{flad2017} and with a Van-Driest damping modifying the model mixing length towards the wall as
\begin{equation}
 l = (C_S \Delta)^2 (1-e^{y^+/A})^m,
\end{equation}
where $|^+$ denotes quantities in wall units and the constants $A$ and $m$ where chosen as $50$ and $3$ respectively, \red{which we found to obtain best results with}. For the filter-based approach the 
filter shape found for DHIT is used, but the 
integration of kinetic energy to determine the filter strength was done only in the wall parallel directions as to obtain a artificial dissipation varying 
with wall distance also within one cell.  The 
constant was set to $c=0.3$, these two measures where found sufficient to obtain proper wall scaling. Note that unlike the Smagorinsky model, where the 
viscosity is computed from the shear strain and thus plateaus towards the wall, the filter strength goes to zero naturally towards the wall with vanishing 
velocities (proportional to $y$).\\
The computational setup consists of a stretched grid in wall normal direction with a bell shape stretching and a ratio of $4$ from smallest to largest cell and
a constant grid size in wall parallel direction with $8$ cells. The dimension are set as in Moser \cite{moser1999} to $[2\pi,2,\pi]$ for $[x,y,z]$, with 
periodic boundaries in wall parallel directions. With $N=7$, grid 
spacings are $\Delta x^+\approx58$, 
$\Delta z^+\approx29$ and $\Delta y^+_{min/max}\approx7/28$, based on an equidistant inner cell point distribution. The resolution is chosen coarser than is 
usually done to obtain visible differences in between the models, as done also eg. by \cite{Hickel2007}. The flow is driven by a constant pressure gradient 
volume source, fixing the friction Reynolds number.
Fig.~\ref{fig:channel} shows the result for the discussed methods. Besides the very coarse resolution, all methods accurately predict the mean velocity profile 
of the flow. The no-model LES slightly under-estimates the mean velocity in the channel center, indicating that it lacks some dissipation. This result shows 
that as discussed above, the overall resolution w.r.t. turbulence resolution is still high. Some differences are seen in the Reynolds stress profiles, mainly 
in $(u'u')^+$. The no-model LES overestimates the Reynold stress peak and underestimates the stress in the center, this is commonly observed for this method 
also eg. in~\cite{wiart2015}. The best prediction of the Reynolds stresses for this very coarse resolution gives the filter based approach. Overall, all 
methods are capable of giving acceptable results for wall resolved LES.

\section{Summary}
\label{sec:conclusion}

Discontinuous Galerkin and related methods pose an attractive compromise between accuracy through local high order approximations and geometric flexibility through the support of unstructured meshes. These properties make them good baseline schemes for LES in non-trivial domains. The recent development of kinetic energy preserving DG schemes provides a robust discretization and thus opens the possibilities of tuning the numerical dissipation (both for surface and volume terms) as a surrogate LES model. In this work, we have presented such an approach based on specifically designed dissipative solution filters. This is further motivated by the potential of the method to be used in very high Reynolds number regimes where the computation of gradients \red{could then potentially} be skipped all together, as indicated by the test cases. The filter kernel and scaling of filter strength are selected based on considerations from turbulence theory, while the filter coefficients themselves are found through a non-linear optimization in which the kinetic energy and dissipation rate for a decaying homogeneous isotropic turbulence serve as a cost function. We show that the optimized kernels perform very well for the decaying homogeneous isotropic turbulence and Taylor-Green-Vortex cases and yield comparable to better results than an explicit Smagorinsky model for all polynomial degrees considered. By scaling the filter strength with the high mode kinetic energy content, the model preserves laminar flows and allows for smooth transition to turbulence. \red{We found that the filter strength is not independent of the flow case. We recommend using the higher values of the filter strength constant established for infinite Reynolds number flow.} For the case of the plane channel flow, the filter procedure observes the near-wall \blue{behavior} through a scaling of the filter strength. \blue{As this modification is cell local and the wall normal direction in a hexahedral cell is easily identified, this method ports easily to more complex geometries including unstructured meshes.} In summary, the presented, optimized filter-based LES approach takes full advantage of the local polynomial spectrum of the high order solution to construct discretization-aware filter kernels. The proposed approach does not require the computation of solution gradients, nor does it introduce additional time step constraints.

\section{Acknowledgement}

The computations were performed within the hpcdg-project at the High Performance Computing Center Stuttgart (HLRS).

\newpage
\appendix

\newpage
\bibliographystyle{acm}
\bibliography{References}

\end{document}